\pdfoutput=1
\newif\ifcomment
\newif\ifdraft
\drafttrue
\newif\ifextra

\documentclass[
	a4paper, 
	10pt, 
	twoside, 
]{LTJournalArticle}
\usepackage[comma,square,numbers,sort&compress]{natbib}
\usepackage{palatino}
\usepackage{eulervm}
\usepackage{xspace}
\usepackage{dcolumn}
\usepackage{bm}
\usepackage{color}
\usepackage{graphicx}
\usepackage{amsmath}
\usepackage{amsfonts}
\usepackage{amssymb}
\usepackage[T1]{fontenc}
\usepackage{mathrsfs}
\usepackage{calrsfs}
\usepackage{physics}
\usepackage{url}
\usepackage{lineno}
\usepackage[utf8]{inputenc}
\usepackage{autobreak}
\usepackage[separate-uncertainty=true,
table-align-uncertainty=true]{siunitx}
\usepackage{array,multirow}
\usepackage{booktabs}
\newcommand{\pp}           {pp\xspace}

\newcommand{\pPb}          {pPb\xspace}
\newcommand{\pA}           {pA\xspace}

\newcommand{\dAu}          {dAu\xspace}
\newcommand{\PbPb}         {PbPb\xspace}

\newcommand{\AuAu}         {AuAu\xspace}

\newcommand{\pt}           {\ensuremath{p_{\rm T}}}

\newcommand{\RAA}          {\ensuremath{R_{\rm AA}}\xspace}
\newcommand{\RpA}          {\ensuremath{R_{\rm pA}}\xspace}
\newcommand{\ROO}          {\ensuremath{R_{\rm OO}}\xspace}
\newcommand{\RNeNe}          {\ensuremath{R_{\rm NeNe}}\xspace}
\newcommand{\RpO}          {\ensuremath{R_{\rm pO}}\xspace}

\newcommand{\snn}          {\ensuremath{\sqrt{s_{\rm NN}}}}

\newcommand{\hrefurl}[1]   {\href{#1}{\url{#1}}}

\newcommand{\Fig}[1]       {Fig.~\ref{#1}}

\newcommand{\Eq}[1]        {Eq.~(\ref{#1})}

\newcommand{\Figure}[1]    {Figure~\ref{#1}}

\newcommand{\com}[1]       {}

\runninghead{A compendium of cold-nuclear matter baseline predictions in light-ion collisions}
\title{A compendium of cold-nuclear matter baseline predictions in light-ion collisions}

\renewcommand{\maketitlehookd}{%
\begin{abstract}
The recent light-ion collision programme at RHIC and the LHC provides a unique opportunity to investigate the onset of quark-gluon plasma formation and parton energy loss in small systems. A quantitative interpretation of emerging jet quenching measurements requires precise control over cold nuclear matter (CNM) effects, which modify hard-process cross sections independently of any hot-medium dynamics. In this work, we present a comprehensive set of perturbative QCD baseline calculations for nuclear modification factors (\RAA) in proton-oxygen (pO), oxygen-oxygen (OO) and neon-neon (NeNe) collisions at LHC energies. The study includes charged hadron, neutral pion, prompt photon, and electroweak-boson production computed at next-to-leading order using a broad set of recent nuclear parton distribution functions (nPDFs).
We demonstrate that CNM effects alone can induce sizeable suppressions in light-ion systems, with large associated nPDF uncertainties that currently limit the quantitative extraction of parton energy loss. To address this limitation, we explore a range of multi-cross-section ratios in which CNM effects and their uncertainties largely cancel. In particular, ratios of neutral pion \ROO to prompt photon \ROO or charged hadron \ROO to $\RpO^2$ provide theoretically robust observables with substantially reduced nPDF uncertainties, thereby enhancing sensitivity to possible energy-loss signatures.
\end{abstract}
}

\author{%
	Florian Jonas\textsuperscript{1}, 
    Constantin Loizides\textsuperscript{2},
    Aleksas Mazeliauskas\textsuperscript{3}, \\ 
    Petja Paakkinen\textsuperscript{1} and
    Nicolas Strangmann\textsuperscript{4}  
}

\date{\footnotesize\textsuperscript{\textbf{1}}CERN, 1211 Geneva 23, Switzerland\\ 
                    \textsuperscript{\textbf{2}}Department of Physics \& Astronomy, Rice University, Houston, 77005 Texas, USA\\
                    \textsuperscript{\textbf{3}}Institute for Theoretical Physics, University of Heidelberg, 69120 Heidelberg, Germany\\ 
                    \textsuperscript{\textbf{4}}Institute for Nuclear Physics, University of Frankfurt, 60438 Frankfurt am Main, Germany}

\begin{document}
\allowdisplaybreaks
\maketitle

\tableofcontents
\section{Introduction}
\vspace{-0.1cm}
At very high temperatures, ordinary matter undergoes a cross-over phase transition to a dense medium of deconfined quarks and gluons~\cite{Busza:2018rrf}.
This new state of matter, the quark-gluon plasma (QGP), is created in ultra-relativistic collisions of lead (Pb) nuclei at the Large Hadron Collider (LHC) and gold (Au) nuclei at the Relativistic Heavy Ion Collider (RHIC), and has been studied with increasing precision over the last two decades~\cite{BRAHMS:2004adc,PHENIX:2004vcz,PHOBOS:2004zne,STAR:2005gfr,ALICE:2022wpn,CMS:2024krd}.

A key signature of QGP formation is the suppression of high-$p_{\rm T}$ hadron and jet yields, commonly referred to as ``jet quenching''~\cite{Cunqueiro:2021wls,Connors:2017ptx,Qin:2015srf,Wang:2025lct}.
Since hadrons and jets originate from the fragmentation of outgoing partons produced in high-$Q^2$ scatterings, they provide a sensitive probe of the interaction between these partons and the created medium.
The short wavelength of the scattered partons enables them to interact with the quarks and gluons of the QGP at the microscopic level, resulting in energy loss to the medium.
This energy loss mechanism makes both the scattered partons and the resulting observable hadrons sensitive probes of the microscopic structure and transport properties of the QGP.
The modification of the production yields of a given observable in heavy-ion collisions (AA) with respect to proton-proton (pp) can be quantified using the nuclear modification factor:
\begin{equation}
    \RAA(\pt) = \frac{\left. \dd[2]{N_{\rm AA}}\middle/\dd{\pt}\dd{y}\right.}{\left\langle N_{\rm coll} \right\rangle \left. \dd[2]{N_{\rm pp}}\middle/\dd{\pt}\dd{y}\right.} ,
    \label{eq:raa_ncoll}
\end{equation}
where $N_{\rm AA}$ and $N_{\rm pp}$ represent the invariant production yields in AA and pp collisions, respectively, and $\langle N_{\rm coll} \rangle$ denotes the average number of binary nucleon-nucleon collisions, determined through Glauber modelling~\cite{dEnterria:2020dwq} to provide the appropriate scaling of the pp reference. 
Pronounced suppression of inclusive charged hadrons \cite{ALICE:2019hno,ATLAS:2015qmb,ATLAS:2022kqu,CMS:2016xef,CMS:2012aa,STAR:2003fka}, identified light-flavour hadrons \cite{ALICE:2019hno,ALICE:2018mdl,PHENIX:2008saf,PHENIX:2012jha}, and both inclusive \cite{ALICE:2015mjv,ATLAS:2014ipv,ATLAS:2018gwx,CMS:2016uxf,CMS:2021vui,STAR:2020xiv} and semi-inclusive jets \cite{STAR:2017hhs,CMS:2024zjn,ATLAS:2023iad,STAR:2023pal,ATLAS:2018dgb,ALICE:2023jye} have been documented at RHIC and the LHC, demonstrating substantial energy loss in central \PbPb and \AuAu collisions.
The QGP formation in \PbPb and \AuAu collisions is additionally supported by the observation of collective flow phenomena \cite{CMS:2011cqy,ALICE:2010suc,ATLAS:2012at}, including the anisotropic flow coefficient $v_{2}$ and higher-order harmonics, as well as strangeness enhancement \cite{ALICE:2013xmt}.

Surprisingly, several signatures traditionally associated with QGP formation, such as anisotropic flow and strangeness enhancement, have also been observed in small systems, including high-multiplicity pp collisions, proton-nucleus collisions, and peripheral nucleus-nucleus collisions~\cite{Loizides:2016tew,Nagle:2018nvi,Grosse-Oetringhaus:2024bwr,Noronha:2024dtq}.
Despite these clear signs of collective effects in small systems, experimental observation of energy loss in these systems remains elusive.
This is partly because event selection based on high activity (e.g.\ using multiplicity and centrality selections) introduces sizeable selection biases as well as modelling uncertainties related to the collision geometry~\cite{Loizides:2017sqq,Park:2025mbt}.

The study of Oxygen-Oxygen (OO) and Neon-Neon (NeNe) collisions provides a unique opportunity to investigate the threshold for QGP formation in small systems~\cite{Citron:2018lsq}.
RHIC delivered the first OO collisions at $\snn = \SI{200}{GeV}$ in spring 2021, followed by the LHC, which provided OO and NeNe collisions at $\snn = \SI{5.36}{TeV}$, as well as proton-oxygen (pO) collisions at $\snn=\SI{9.62}{TeV}$ in July 2025.
In February 2026, RHIC concluded its operation with a final data taking campaign of OO collisions.
At the time of writing, only several months have elapsed since these data were collected and experimental analyses are ongoing.
Nevertheless, several experimental results are already available:
Initial results from the CMS collaboration \cite{CMSOO} demonstrate significant suppression of unidentified hadron yields in minimum-bias OO collisions at $\snn = \SI{5.36}{TeV}$ relative to the pp baseline, which is supported by preliminary results for $\pi^0$ production from the ALICE collaboration.
These measurements can be described rather well by a wide set of theoretical models that account for partonic energy loss in OO collisions~\cite{Huss:2020whe,Liu:2021izt,Zakharov:2021uza,Ke:2022gkq,Behera:2023oxe,Xie:2022fak,Faraday:2025pto,vanderSchee:2025hoe,Pablos:2025cli}.
Additionally, the ALICE, ATLAS and CMS collaborations report substantial anisotropic flow of charged hadrons in OO and NeNe collisions~\cite{ALICEFlowOO,ATLAS:2025nnt,CMS:2025tga}, providing further evidence for collective behaviour in these small systems~\cite{Giacalone:2024luz,Mantysaari:2025kls}.

To unambiguously attribute the observed yield suppression to jet quenching, precise understanding of all QGP-unrelated effects that may lead to suppression of hadron production in OO collisions is essential.
In particular, initial-state effects unrelated to the hot medium may modify the parton structure in the nucleons of the colliding nuclei, leading to modifications of the production cross sections for various probes produced in high-$Q^2$ processes~\cite{Arleo:2025oos}.
The parton structure of nucleons in the colliding projectiles is commonly described by nuclear parton distribution functions (nPDFs) \cite{Klasen:2023uqj}, which encode the non-perturbative physics of the initial state and are determined through fits to experimental data.
The modifications due to these cold-nuclear matter (CNM) effects can be substantial, and include gluon shadowing at low $x$, anti-shadowing, as well as Fermi motion at the largest $x$ values.
Furthermore, isospin effects, arising from the ratio of protons and neutrons in the colliding nuclei, may influence the observed production yields.
Consequently, measurements of nuclear modification factors must be compared with CNM baseline calculations to draw meaningful conclusions about the presence and magnitude of jet quenching in light-ion collisions.

In minimum-bias collisions, the perturbative QCD (pQCD) baseline for nuclear modification factor can be computed  as the ratio of cross sections
\begin{equation}
    R_{\rm{AB}}^\text{min.-bias}(\pt) = \frac{1}{AB}\frac{\left. \dd[2]{\sigma_{\rm AB}}\middle/\dd{\pt}\dd{y}\right.}{\left. \dd[2]{\sigma_{\rm pp}}\middle/\dd{\pt}\dd{y}\right.},
    \label{eq:raa_minbias}
\end{equation}
where $A$ and $B$ are the atomic numbers of the colliding nuclei ($A=B=16$ for OO, $A=B=20$ for NeNe and $A=1, B=16$ for pO collisions). $\sigma_{\rm AB}$ is a cross section for a hard process in (potentially asymmetric) AB nucleus collisions, which can be systematically computed order-by-order using QCD factorization~\cite{Collins:1989gx,Metz:2016swz}.

\begin{figure*}[t]
    \centering
    \includegraphics[width=\linewidth]{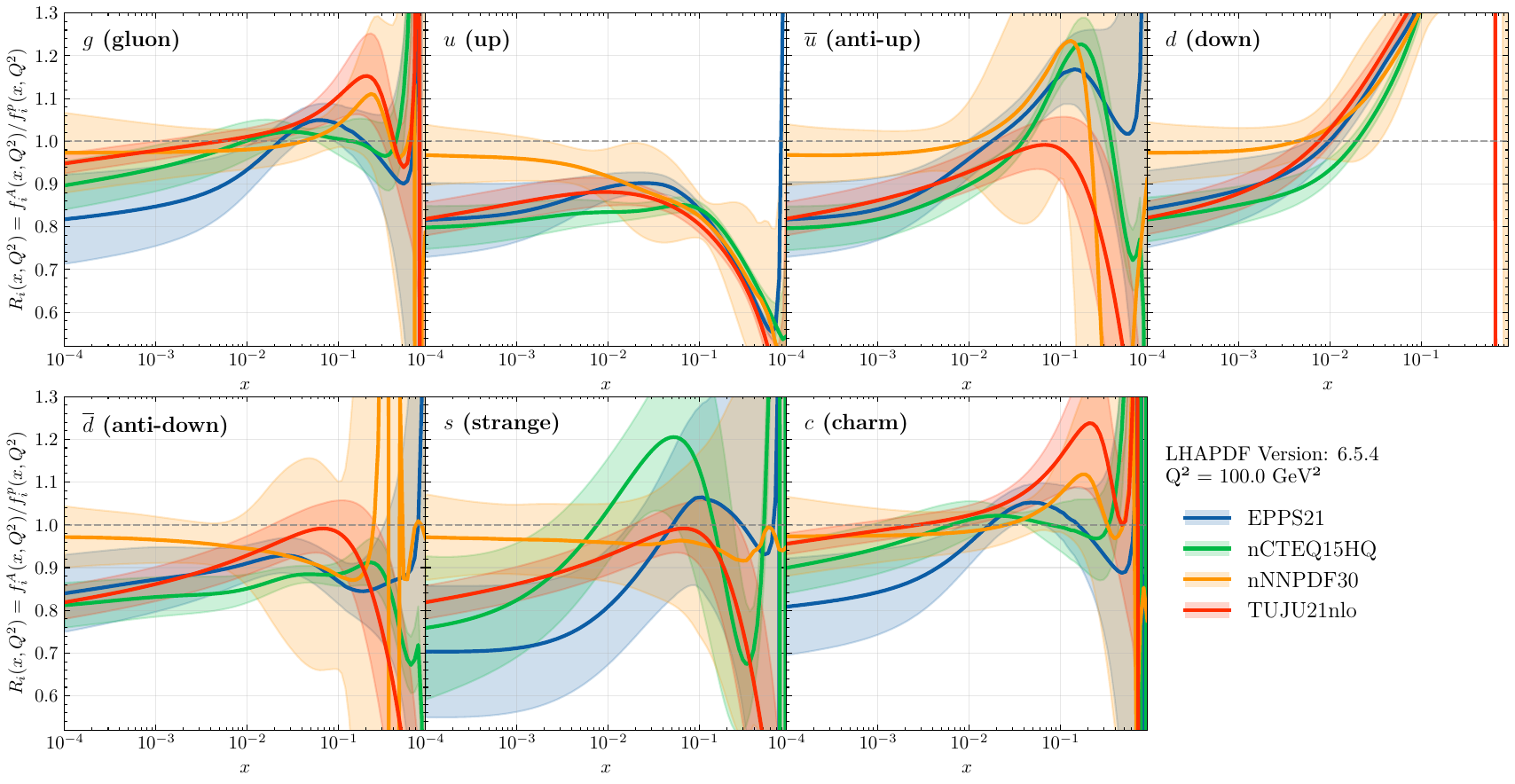}
    \caption{$x$ dependence of nPDFs for different partons for oxygen nucleus at $Q^2=\SI{100}{GeV}^2$ with respect to the corresponding free-proton baseline. The bands denote the \SI{68}{\percent} nPDF confidence interval.}
    \label{fig:npdfsvsxQ2100GeV}
\end{figure*}

The purpose of this article is twofold: First, we provide a comprehensive set of baseline pQCD calculations for nuclear modification factors \Eq{eq:raa_minbias} in pO, OO and NeNe collisions that include CNM effects but exclude QGP effects such as jet quenching. Our study complements previous works~\cite{Huss:2020dwe,Brewer:2021tyv,Belmont:2023fau,Gebhard:2024flv,Mazeliauskas:2025clt,Faraday:2025prr,Fuks:2024ctk} with computations using a comprehensive list of nPDFs from recent global analyses and a wide selection of hadronic and electroweak observables.
Here we concentrate on fully inclusive observables; for discussion on semi-inclusive, jet- and hadron triggered observables, see Ref.~\cite{Gebhard:2024flv}.
Second, given the sizeable nPDF uncertainties in nuclear modification factors, we study multi-cross-section ratios that could potentially cancel CNM effects and provide more precise baselines for jet quenching signatures.
In this paper, our main focus is on predictions at central rapidities and LHC energies, but the calculations can be readily extended to different kinematic cuts and collision energies.
We aim to establish a comprehensive set of CNM baselines to support ongoing experimental investigations of energy loss in light-ion collisions and to motivate dedicated analyses of new energy-loss observables.

This article is structured as follows:
Section \ref{sec:lhapdf} outlines the current landscape of nuclear PDFs and discusses their $A$ dependence.
Section \ref{sec:hadronproduction} provides calculations for the nuclear modification factors of charged hadrons ($h^\pm$) and neutral pions ($\pi^0$), including the centre-of-mass energy dependence of their production.
Section \ref{sec:ewbosonproduction} presents the nuclear modification factors for $W^{\pm}$, $Z$ bosons and prompt photons.
Finally, in Sections~\ref{sec:hadron_over_ewboson}, \ref{sec:soo} and \ref{sec:rnene_roo} we discuss a variety of multi-cross-section ratios that allow for the cancellation of nPDF uncertainties, which may be used for searches of QGP-like effects in OO and NeNe collisions. We conclude in Section \ref{sec:summary}.

All the results presented in this work are publicly available~\cite{GitLabCNMData}.

\section{Recent nuclear PDFs}
\label{sec:lhapdf}
Nuclear PDFs encode the partonic structure of nuclei at high energy and corresponding CNM effects, assuming a factorization of the long and short length scales of the interaction and the universality of the underlying parton distributions.
In particular, nPDFs are commonly given as a set of functions $f_i^A(x,Q^2)$, which at leading order (LO) in pQCD can be interpreted as the probability densities for finding a parton $i$ that carries a fraction $x$ of the momentum of a nucleon that is embedded in a nucleus with mass number $A$ at a given scale $Q^2$.
While the $Q^2$ dependence can be calculated perturbatively in pQCD, the $x$-dependence must be constrained using experimental data.
The extraction of these nPDFs from experimental data has been performed in various global analyses, where the choice of input data, the functional ansatz for $f_i^A(x,Q^2)$, the treatment of experimental and theoretical uncertainties, and the fitting procedure differ significantly between analyses.
For a comprehensive overview of currently available nPDFs and their differences, we refer the reader to a recent review by Klasen and Paukkunen given in Ref.~\cite{Klasen:2023uqj}.

\begin{figure*}[t]
    \centering
    \includegraphics[width=0.49\linewidth]{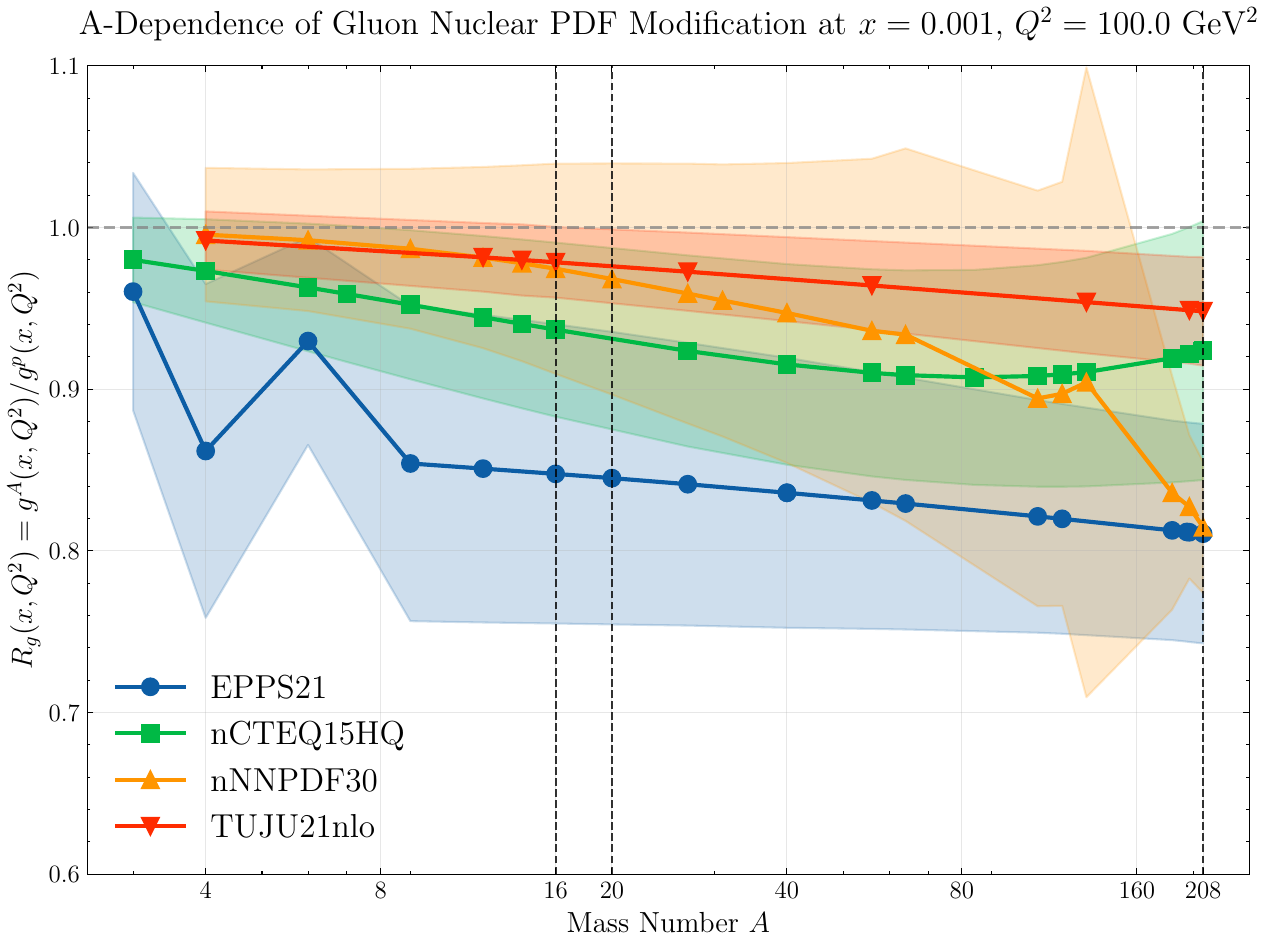}
    \includegraphics[width=0.49\linewidth]{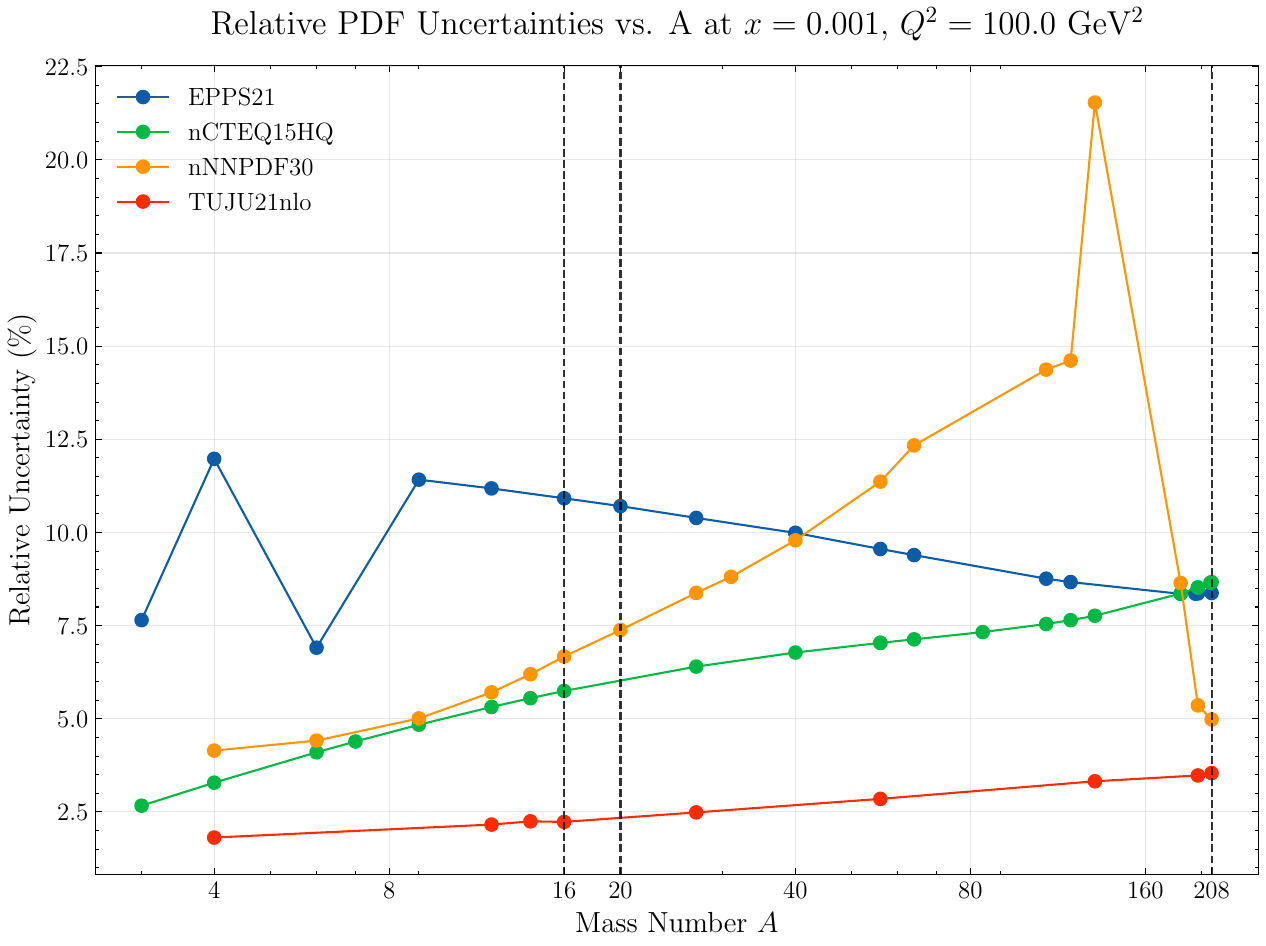}
    \caption{(Left) Modification of the gluon nPDF with respect to the corresponding free-proton baseline at $x=10^{-3}$ and $Q^2=\SI{100}{GeV}^2$ as a function of mass number $A$. Bands show 68\% confidence intervals. (Right) Relative nPDF uncertainty as a function of mass number $A$.}
    \label{fig:AdependencenPDFs}
\end{figure*}

The current state of CNM effects in oxygen ($A=16$) is summarized in Fig. \ref{fig:npdfsvsxQ2100GeV}, which shows the $x$ dependence of recent nuclear PDFs for oxygen at an exemplary $Q^2$ of $\SI{100}{GeV}^2$ for various parton flavours.
In particular, for each parton flavour $i$ the ratio:
\begin{equation}
    R_i^A(x, Q^2) = \frac{f_i^A(x, Q^2)}{f_i^{p}(x, Q^2)}
    \label{eq:npdf}
\end{equation}
is shown, which corresponds to the modification of the nPDF with respect to the corresponding free-proton baseline.
Because of the ratio with free proton PDF, these plots for up and down flavours reflect also the isospin effects in addition to the nontrivial nuclear modifications.
The parton distributions have been accessed using the LHAPDF6 library \cite{lhapdf6}, which has also been used to calculate the nPDF uncertainties at $\SI{68}{\percent}$ confidence level, denoted by shaded bands. Note that nuclear PDF error sets often include the uncertainty from the corresponding free proton baseline, which can be cancelled in ratios such as Eqs.~\eqref{eq:raa_minbias} and \eqref{eq:npdf}. Because proton PDF uncertainties are much smaller than nPDF uncertainties, this typically results in only a sub-percent reduction in the uncertainty bands. For computational efficiency, we will not cancel proton PDF uncertainties except for the results shown in Section~\ref{sec:soo}.

To date, in their latest versions, four global analyses provide nPDFs for oxygen that are available in LHAPDF6: EPPS21 \cite{Eskola:2021nhw}, nCTEQ15HQ \cite{nCTEQ15HQ}, nNNPDF30 \cite{nNNPDF30} and TUJU21 \cite{TUJU21}.
None of these global fits use oxygen data directly; instead, the oxygen nPDFs are obtained by interpolating between the free proton baseline and data from other moderate-to-heavy nuclei. 
This is typically done by including an assumed functional form of the $A$ dependence as part of the fitting procedure.
All available nPDFs show sizeable modifications in oxygen with respect to the free-proton baseline; however, both the uncertainties and the magnitude of the modifications vary significantly among the different global analyses.
For example, while EPPS21 shows gluon shadowing in oxygen at $x\sim10^{-4}$ of about \SI{20}{\percent}, smaller modifications of less than \SI{10}{\percent} are observed for the other three analyses.

Although significant progress has been made in constraining nPDFs in the last decade thanks to the heavy-ion program at RHIC and the LHC, Fig. \ref{fig:npdfsvsxQ2100GeV} highlights that there is still much to learn about CNM effects in nuclei.
In particular, the $A$-dependence of nuclear modifications for many flavour combinations is only poorly constrained, as inputs to current nPDF fits are dominated by a variety of Deep Inelastic Scattering (DIS) data from light-to-heavy nuclei, but only \pPb data at the LHC.
Currently, global nPDF analyses lack collider data for light nuclei, which are critical for constraining the small-$x$ regime. 
Furthermore, the measurements of the $A$ dependence in DIS experiments, e.g.\ at SLAC \cite{Arnold:1983mw} and NMC \cite{NewMuon:1996yuf} can be fitted with various parametric forms, not being able to offer an unambiguous insight on how the nPDFs should be parametrised as a function of the mass number in general.

\begin{table*}
    \centering
    \begin{tabular}{c|lll}
    \toprule 
        &Name & LHAPDF name &  Proton reference \\
        \midrule
        \parbox[t]{2mm}{\multirow{4}{*}{\rotatebox[origin=c]{90}{Oxygen}}} &EPPS21 \cite{Eskola:2021nhw} & \texttt{EPPS21nlo\_CT18Anlo\_O16}  & \texttt{CT18ANLO}\\   
        &nCTEQ15HQ \cite{nCTEQ15HQ} & \texttt{nCTEQ15HQ\_FullNuc\_16\_8} & \texttt{CT18NLO}\\
        &nNNPDF30 \cite{nNNPDF30}& \texttt{nNNPDF30\_nlo\_as\_0118\_A16\_Z8} & \texttt{nNNPDF30\_nlo\_as\_0118\_p} \\
        &TUJU21nlo \cite{TUJU21} & \texttt{TUJU21\_nlo\_16\_8} & \texttt{TUJU21\_nlo\_1\_1}\\
       \midrule
       \parbox[t]{2mm}{\multirow{2}{*}{\rotatebox[origin=c]{90}{Neon}}} &EPPS21 \cite{Eskola:2021nhw}& \texttt{EPPS21nlo\_CT18Anlo\_Ne20} & \texttt{CT18ANLO}\\  
        &nNNPDF30 \cite{nNNPDF30}& \texttt{nNNPDF30\_nlo\_as\_0118\_A20\_Z10} & \texttt{nNNPDF30\_nlo\_as\_0118\_p} \\
       \bottomrule
    \end{tabular}
    \caption{Overview of nPDFs used in this work, including their name in the LHAPDF6 \cite{lhapdf6} library\protect\footnotemark.
    In addition, for each nPDF the corresponding PDF used to calculate the pp reference is given.}
    \label{tab:pdftable}
\end{table*}

To illustrate this point, Fig.~\ref{fig:AdependencenPDFs} (left) shows the $A$-dependence of the gluon modification $R_g^A(x, Q^2)$ at $x=0.001$ and $Q^2=\SI{100}{GeV}^2$ for the considered nPDFs.
The shaded band denotes the nPDF uncertainty at $\SI{68}{\percent}$ confidence level, which is for clarity also given as a relative uncertainty in Fig.~\ref{fig:AdependencenPDFs} (right).
For all nPDFs, the amount of gluon shadowing tends to increase with increasing $A$, but the way how this happens varies a lot between different parametrisations.
For EPPS21, nCTEQ15HQ and TUJU21, the $A$-dependence is introduced in the global fit via an analytical parametrisation (either modifying the free fit parameters directly or constraining the nuclear to free-proton ratio), leading to the power-law-like increase of the suppression observed in Fig.~\ref{fig:AdependencenPDFs}.
As these nuclear degrees of freedom are introduced at the starting scale, the observed $A$ dependence may not be strictly monotonic at a given scale $Q$, leading, e.g.\ to the slight rise for high $A$ observed for nCTEQ15HQ.
For EPPS21, the rate of change in gluon shadowing as a function of $A$ is rather modest, except at very light nuclei, where helium-3 and lithium-6 deviate from the general trend. This happens due to additional parameters included in the fit, which were introduced to better accommodate high-$x$ DIS data for these nuclei~\cite{Eskola:2021nhw}. The presence of this non-monotonic behaviour for the lightest nuclei in EPPS21 but not in any other fit emphasises the significant parametrisation dependence and lack of strong experimental data controls for light-nuclei gluons.
For nNNPDF30, the $A$ dependence is introduced not analytically but in the neural network component of the fit, which may explain the more linear $A$ dependence and slight fluctuations as a function of $A$.
The $A$-dependence of the nPDF uncertainties for gluons at low-$x$ (Fig.~\ref{fig:AdependencenPDFs} right) likewise shows significant differences between different global analyses, both in magnitude and $A$-dependence.

These large variations in the light-nuclei parton distribution functions have become of high phenomenological relevance since they represent the primary source of 
uncertainty in the hadron nuclear modification factor baseline for light-ion collisions~\cite{Huss:2020dwe,Brewer:2021tyv,Belmont:2023fau,Gebhard:2024flv,Mazeliauskas:2025clt,Faraday:2025prr}. 
As a result, up to half of the observed suppression in minimum-bias hadron \ROO at $\snn = \SI{5.36}{TeV}$ can be explained by nPDF effects~\cite{CMSOO}, but with a very large uncertainty that prohibits making precise quantitative claims about the magnitude of jet quenching effects in these systems.

The previous considerations highlight the need for LHC data at light-ions such as oxygen and neon to better constrain the $A$-dependence of CNM effects at low-$x$ and high $Q^2$.
The 2025 LHC pO run provides a unique dataset to bridge this gap. Integrating these results, for example, on 
di-jet production, into 
future global fits will significantly reduce PDF uncertainties and improve the 
precision of small-$x$ distributions, particularly for the gluon density $g_A(x, Q^2)$~\cite{Paakkinen:2021jjp}. 
Similarly to di-jets, the inclusive hadron production (see Sec.~\ref{sec:hadronproduction}) can be measured with good statistics even with relatively low luminosities, and the pO data can be expected to yield new constraints. Towards low $p_{\rm T}$, the hadron production can be subject to additional CNM effects beyond nPDFs~\cite{Arleo:2025oos}, and therefore it is important to test the universality of the obtained constraints with additional observables. Such a test is possible with the production of electroweak bosons (see Sec.~\ref{sec:ewbosonproduction}). Due to the electroweak couplings involved in these processes they are more luminosity hungry, but thanks to the very successful 2025 LHC light-ion campaign, their measurement in pO and OO should be feasible and yield new insight on the parton distribution nuclear modifications.

\begin{figure*}[t]
    \centering
    \includegraphics[width=0.33\linewidth]{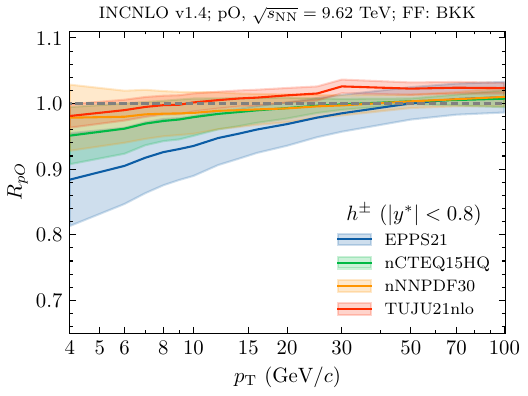}
    \includegraphics[width=0.33\linewidth]{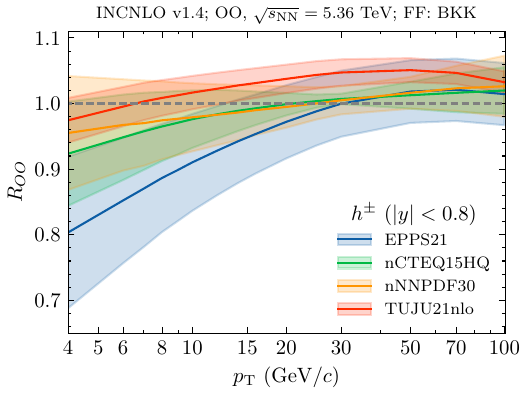}
     \includegraphics[width=0.33\linewidth]{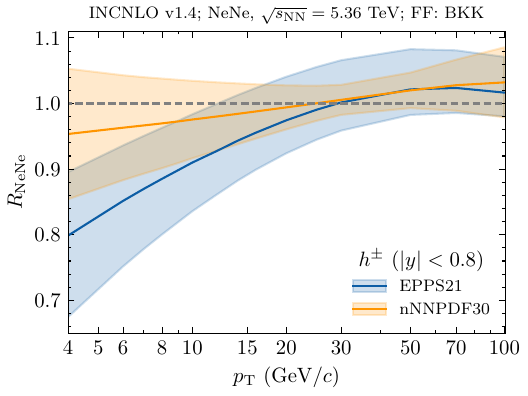}
    \caption{Nuclear modification factors \RpO (left), \ROO (middle) and \RNeNe (right) of charged hadron production. The shown uncertainty bands denote the uncertainty of the nPDF at \SI{68}{\percent} confidence level.}
    \label{fig:hadronROOandRpO}
\end{figure*}
\footnotetext{nPDFs for neon were obtained from  \url{https://research.hip.fi/qcdtheory/nuclear-pdfs/epps21/} for EPPS21 and by private communication for nNNPDF30.}
\section{Hadron production}
\label{sec:hadronproduction}
Hadrons with large transverse momenta originate from the fragmentation of high-virtuality outgoing partons. These partons evolve through a parton shower and subsequent hadronization into observable particles, making hadron production a sensitive probe of both energy loss and CNM effects.
Measurements of charged hadrons in \PbPb collisions \cite{ALICE:2019hno,ATLAS:2015qmb,ATLAS:2022kqu,CMS:2016xef,CMS:2012aa} are significantly suppressed with respect to the pp baseline, compatible with parton energy loss in the QGP.
While no energy loss has been observed in \pPb collisions, charged hadron \cite{ALICE:2014nqx,ALICE:2018vuu,CMS:2016xef} and $\pi^0$ production \cite{ALICE:2021est,LHCb:2022tjh,ALICE:2018vhm} are significantly suppressed at low-$p_{\rm T}$ due to gluon shadowing, well described by recent nPDFs.
First measurements by the CMS collaboration \cite{CMSOO} demonstrate significant suppression of unidentified hadron yields in minimum-bias OO collisions at $\snn = \SI{5.36}{TeV}$. 

In the following, we provide no-quenching baselines for the nuclear modification factors \RpO, \ROO and \RNeNe  for charged hadrons ($h^{\pm}$) and neutral pions ($\pi^0$).
We provide predictions for neutral pions, as well as charged hadrons in pO collisions at $\snn = \SI{9.62}{TeV}$.
The calculations are performed at NLO using the INCNLOv1.4 program \cite{incnlowebsite,INCNLO1,INCNLO2}.
The fragmentation of an outgoing parton to a final-state hadron is described using the BKK fragmentation function \cite{BKK}.
Sensitivity of \RAA to fragmentation functions was studied in Ref.~\cite{Mazeliauskas:2025clt} and was found largely negligible.
The resulting hadron is required to be produced at mid-rapidity ($|y^*|<0.8$) in the centre-of-mass frame. We will denote the lab-frame rapidity as $y$ which coincides with the centre-of-mass rapidity $y^*$ for symmetric collision systems in the collider mode at the LHC.
The renormalisation scale $\mu_{\rm R}$, factorization scale $\mu_{\rm F}$ and fragmentation scale $\mu_{\rm FF}$ are chosen to coincide with the transverse momentum ($p_{\rm T}$) of the hadron.

For pO collisions and OO collisions, the calculation is performed at a centre-of-mass energy per nucleon pair of $\sqrt{s_{\rm NN}}=\SI{9.62}{TeV}$ and $\sqrt{s_{\rm NN}}=\SI{5.36}{TeV}$, respectively, which corresponds to the collision energy during LHC operation for the light-ion data taking in the summer of 2025.
While the proton reference data taking campaign was only performed at $\sqrt{s_{\rm NN}}=\SI{5.36}{TeV}$, we choose for each calculation the same centre-of-mass energy in pp collisions as for the light-ion collisions, as measurements of the \RpA and \RAA commonly use extrapolations for the pp reference for the cases of non-coinciding centre-of-mass energies.
The used nPDFs are denoted in the respective legends in the following figures, and the respective free-proton reference PDF is given in Tab.~\ref{tab:pdftable}.

Figure~\ref{fig:hadronROOandRpO} shows the charged hadron nuclear modification factors for proton-oxygen (left), oxygen-oxygen  (middle) and neon-neon (right).
A suppression of hadron production of up to \SI{10}{\percent} (\SI{15}{\percent}) is observed for the \RpO (\ROO) at $p_{\rm T}\sim\SI{6}{GeV}/c$ for EPPS21, which can be attributed to gluon shadowing, as discussed in Sec.~\ref{sec:lhapdf}.
Less suppression at low $p_{\rm T}$ is observed for the other studied nPDFs, however, all predictions agree within the sizeable nPDF uncertainties denoted by shaded bands at \SI{68}{\percent} confidence level.
The scale uncertainty was found to be negligible and is not shown.
As expected, the observed modifications are larger for OO collisions than pO collisions, and we observe in particular $\ROO\approx \RpO^2$ for CNM effects.
Figure~\ref{fig:hadronROOandRpO}~(right) shows the nuclear modification factor \RNeNe for neon projectiles for EPPS21 and nNNPDF30, which are the only collaborations that provide neon grids at the time of writing this manuscript.
The observed modification due to CNM effects is very similar to the \ROO, in particular one observes an at most \SI{1}{\percent} increased suppression of the central value at low $p_{\rm T}$.
The ratio $\RNeNe/\ROO$ is further discussed in Sec. \ref{sec:rnene_roo}, including the correlation of nPDF uncertainties.

We have also computed the neutral pion nuclear modification factors. We found that for $\pi^0$, the nuclear modification factors agree within \SI{1}{\percent} with those of charged hadron production. We include these calculations in the publicly available repository~\cite{GitLabCNMData}.

\begin{figure}[t]
    \centering
     \includegraphics[width=\linewidth]{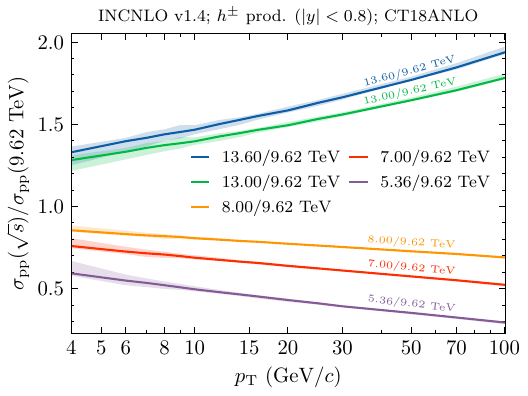}
    \caption{Cross-section ratio of charged hadron production in pp collisions at various centre-of-mass energies with respect to the cross section at $\sqrt{s_{\rm NN}}=\SI{9.62}{TeV}$. Scale uncertainties are denoted as shaded bands.}
    \label{fig:cmedependence}
\end{figure}

\begin{figure}[t]
    \centering
     \includegraphics[width=\linewidth]{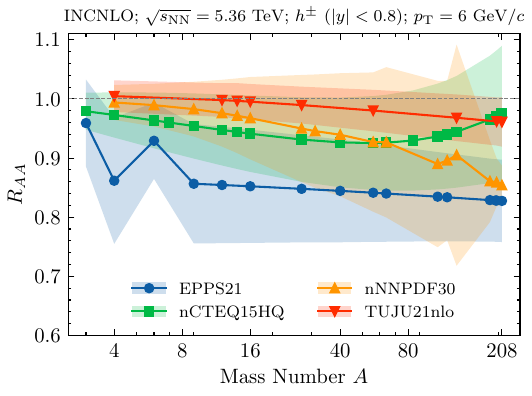}
    \caption{Mass number ($A$) dependence of the nuclear modification factor \RAA of charged hadron production at $p_{\rm T}=\SI{6}{GeV}/c$. The uncertainty band denotes the nPDF uncertainties at \SI{68}{\percent} confidence level.}
    \label{fig:Adependencepi0}
\end{figure}

Experimental measurements of the \RpO require a pp reference at $\sqrt{s_{\rm NN}}=\SI{9.62}{TeV}$ obtained from interpolations, as there is no LHC dataset available at this centre-of-mass energy.
To aid these interpolations, we provide in Fig. \ref{fig:cmedependence} various charged hadron cross-section ratios of all available centre-of-mass energies for pp collisions at the LHC with respect to $\sqrt{s}=\SI{9.62}{TeV}$.
We found that while the proton PDF uncertainties cancel almost completely and are negligible on the ratios, the scale uncertainties do not fully cancel, and are therefore denoted as shaded bands in the figure.
To estimate the scale uncertainty, the three scales $\mu_{\rm R}$,  $\mu_{\rm F}$ and $\mu_{\rm FF}$ are varied individually from $0.5\pt$ to $2\pt$, resulting in $15$ variations after excluding the extreme cases where a scale of $2\pt$ is paired with a scale of $0.5\pt$.
All scales are varied simultaneously at both respective collision energies in order to account for correlations of the scale uncertainties.
Scale uncertainties of up to \SI{10}{\percent} are observed at $p_{\rm T}\sim\SI{4}{GeV}/c$, reducing to less than \SI{5}{\percent} at high $p_{\rm T}$.

Finally, we present the dependence of the charged hadron nuclear modification factor on the mass number $A$ of the colliding nucleus in Fig. \ref{fig:Adependencepi0}, including the corresponding sizeable nPDF uncertainties of the respective sets.
The modification factor is shown at an exemplary \pt\ of $\SI{6}{GeV}/c$, which probes the gluon shadowing regime.
All features discussed on the level of nPDFs in Fig. \ref{fig:AdependencenPDFs} clearly propagate to the charged hadron modification factor, highlighting the resulting sizeable uncertainties of pQCD calculations for light ions.

\begin{figure*}[h!]
    \centering
    \includegraphics[width=1.0\linewidth]{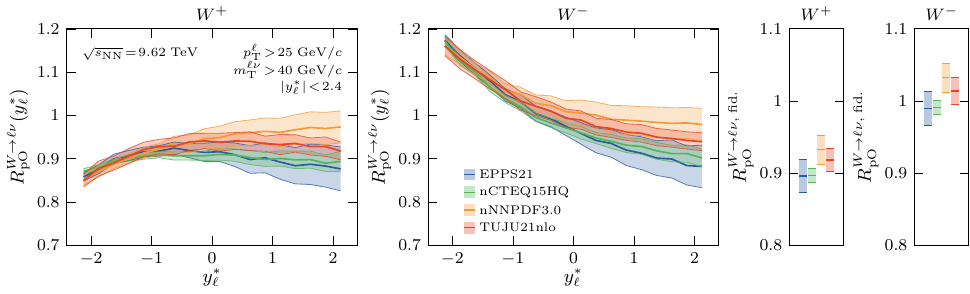}\\
    \includegraphics[width=1.0\linewidth]{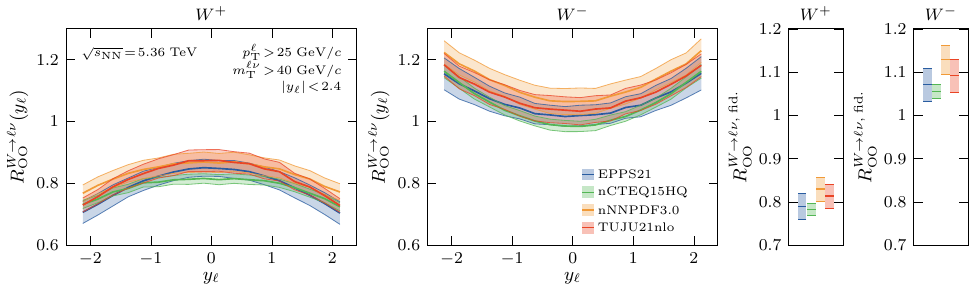}
    \caption{Lepton-rapidity differential and fiducial inclusive \RpO (top) and \ROO (bottom) for $W^\pm$ production. The calculations are performed at NLO with the MCFM code~\cite{MCFM}. The shown uncertainty bands denote the  nPDF uncertainties at \SI{68}{\percent} confidence level.}
    \label{fig:WBosonRatios}
\end{figure*}

\section{Electroweak-boson production}
\label{sec:ewbosonproduction}
Electroweak-boson production serves as an important benchmark process at hadron colliders and their production in \pp and \pPb collisions have been routinely studied in analyses of proton and nuclear PDFs.
Since electroweak bosons do not carry colour charge, studying their production in heavy-ion collisions is particularly interesting, as they do not lose energy in a hot medium and modification of their production is dominated by CNM effects.
This makes them an important calibration probe for OO and NeNe collisions to disentangle CNM effects from potential hot medium effects observed for other observables.

\subsection{$W^\pm$ and $Z$ boson production}

Measurements of $W^\pm$ and $Z$ bosons in \PbPb~\cite{ATLAS:2019ibd,ATLAS:2019maq,CMS:2021kvd,ALICE:2022cxs} and \pPb~\cite{LHCb:2014jgh,ATLAS:2015mwq,CMS:2015zlj,CMS:2015ehw,ALICE:2016rzo,CMS:2019leu,ALICE:2020jff,CMS:2021ynu,LHCb:2022kph,ALICE:2022cxs} collisions show significant modifications with respect to no CNM effect predictions and have provided important constraints for nPDF analyses.
$W^\pm$ bosons are at leading order dominantly produced by an interaction of a valence quark and a sea quark, in particular through $\text{u}\bar{\text{d}}\rightarrow W^+$ and $\text{d}\bar{\text{u}}\rightarrow W^-$ channels, but giving also sensitivity to the strange quarks through subleading contributions~\cite{Kusina:2020lyz} and gluons via scale evolution and higher-order terms.
The production rates of $W^\pm$ bosons in ion collisions are modified with respect to a free-proton baseline due to isospin effects as well as a modification of the PDFs themselves.
The former arise as a nucleus containing protons as well as neutrons naturally has a different average valence quark distribution than a single proton.
The latter may arise e.g.\ due to the shadowing of quarks and gluons at low-$x$ in the involved nucleus, as illustrated in Fig.~\ref{fig:npdfsvsxQ2100GeV}.
The production of $Z$ bosons is likewise affected by CNM effects such as shadowing but not significantly by isospin effects, in contrast to $W^\pm$ production.
The absolute cross sections for the $W^\pm$ and $Z$ production can show a non-negligible dependence on proton PDFs in comparison with the nuclear modification uncertainties~\cite{Paukkunen:2010qg,Eskola:2022rlm}, and thus we are considering here only the nuclear modification factors, where the proton-PDF dependence can be shown to cancel to a good extent~\cite{Eskola:2022rlm}.

In the following, we provide predictions for the expected modification of the production rates of $W^\pm$ and $Z$ bosons in OO and pO collisions.
The calculations are performed at NLO using the MCFM 10.1~\cite{MCFM} program for the semi-leptonic decays $W^\pm\rightarrow l^\pm \nu$ and $Z\rightarrow l^\pm l^\mp$.%
\footnote{In particular, we use a modified version of MCFM10.1 used in \cite{Jonas:2021xju} that allows specifying the (n)PDF for each beam separately. This allows us also to obtain the results for pO collisions.}
The calculations for $W^\pm$ production are performed in the exemplary fiducial acceptance $p_{\rm T}^l> \SI{25}{GeV}/c$, $m_{\rm T}^{l\nu}>\SI{40}{GeV}/c^2$ and $|y_l|<2.4$, which is commonly used in measurements by the CMS collaboration~\cite{CMS:2024myi}. We use the same rapidity interval in the centre-of-mass frame for pO, which due to the boost of the system would correspond to a rapidity interval of approximately $-2.05<y_l<2.75$ in the laboratory frame, thus exceeding the CMS muon acceptance. However, we have checked that the impact of restricting the rapidity to a narrower $|y^*| < 2.05$ interval is a marginal, percent level effect on the ratios of fiducial inclusive cross sections. For $Z$ production, a lower cut for the lepton transverse-momentum $p_{\rm T}^l> \SI{15}{GeV}/c$ is used together with a dilepton invariant mass requirement of $\SI{60}{GeV}/c^2<m_{ll}<\SI{120}{GeV}/c^2$. The factorisation and renormalisation scales are fixed to $\mu_F = \mu_R = m_{\rm T}^{l\nu}$ for the $W^\pm$ and $\mu_F = \mu_R = m_{ll}$ for the $Z$ production, but the results for the nuclear modification factors are expected to be insensitive to this choice and variations around the central scale.

Figure \ref{fig:WBosonRatios} (top row) shows the nuclear modification factor \RpO for $W^+$ and $W^-$ production as a function of lepton rapidity in the centre-of-mass frame $y^*_l$ (first and second panel) at $\sqrt{s_{\rm NN}}=\SI{9.62}{TeV}$, with the pp baseline being expected to be interpolated to this energy. For negative rapidities, i.e.\ the oxygen-going direction, where one probes dominantly the nuclear valence quarks, the predictions obtained with different nPDFs agree well. This is due to the valence-quark distributions in these nPDFs being constrained for oxygen through $A$-interpolation from fixed-target DIS data for other nuclei. The differences between the $W^+$ and $W^-$ ratios in this negative-rapidity region originate predominantly from isospin effects. The $W^-$ boson, coupling preferentially to down valence quarks of the nucleus, receives an enhancement from the presence of neutrons, which increase the down-quark distribution with respect to the proton reference, as can be seen also in Fig. \ref{fig:npdfsvsxQ2100GeV}. Conversely, the $W^+$ boson gets a suppression from the relative lower number of protons and thus up quarks in the nucleus. As the lepton rapidity increases to positive values, i.e.\ to the proton-going direction, the predictions from different nPDFs begin to deviate from each other due to differences in the small-$x$ sea-quark and gluon distributions. This results in a total uncertainty envelope of approximately $\pm\SI{10}{\percent}$ at $y^*_l \approx 2$, with nNNPDF3.0 giving the largest and EPPS21 the smallest values for the ratio. With the produced $W^\pm$ bosons originating in these rapidities preferentially from valence quarks of the proton both in the pO and pp collisions, the isospin effects become significantly smaller. The observed differences are also reflected in the fiducial inclusive ratios shown also in Fig. \ref{fig:WBosonRatios} (top row, third and fourth panel), with the isospin effects explaining why $\RpO^{W^+}<\RpO^{W^-}$ and the variations in different nPDF predictions originating from different amounts of shadowing in small-$x$ quark and gluon distributions. For these inclusive ratios the nPDF uncertainties are somewhat diminished due to integrating over the regions of negative and positive rapidities.

The results for $W^\pm$ production in OO at $\sqrt{s_{\rm NN}}=\SI{5.36}{TeV}$ are shown in Fig. \ref{fig:WBosonRatios} (bottom row) in a similar fashion to the pO results above. Now, since the quark coupling to the $W^\pm$ comes always from the nucleus, we find strong isospin effects throughout the rapidity range, leading to $\ROO^{W^+}<\ROO^{W^-}$ also for the fiducial inclusive ratios. Similarly, since the small-$x$ nuclear modifications are probed (symmetrically) both in the positive and negative directions, we see a separation in the predictions at all rapidities, with EPPS21 and nCTEQ15HQ giving more suppressed ratios compared to nNNPDF3.0 and TUJU21. Due to the smaller collision energy, and thus more limited small-$x$ reach, this separation is less strong compared to the pO results. Apart from the isospin effects, the difference between $W^+$ and $W^-$ ratios depends on the nontrivial nuclear modifications. In fact, since oxygen is isoscalar, the leading contributions $\text{u}\bar{\text{d}}\rightarrow W^+$ and $\text{d}\bar{\text{u}}\rightarrow W^-$ are identical in OO (assuming isospin symmetry between the bound proton and neutron PDFs), and the difference of $W^+$ and $W^-$ cross sections is then directly sensitive to the subleading Cabibbo-suppressed contributions.  A precise enough measurement of the $W^\pm$ cross sections or nuclear modification factors could therefore help constraining the currently rather poorly understood strange-quark nPDFs.

\begin{figure}[t]
    \centering
    \includegraphics[width=1.0\linewidth]{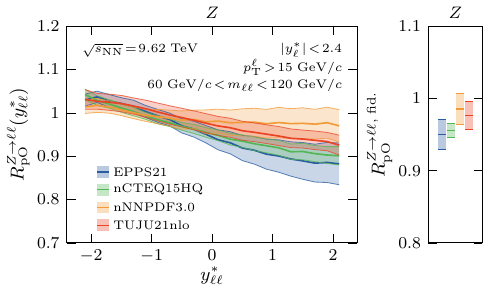}
    \includegraphics[width=1.0\linewidth]{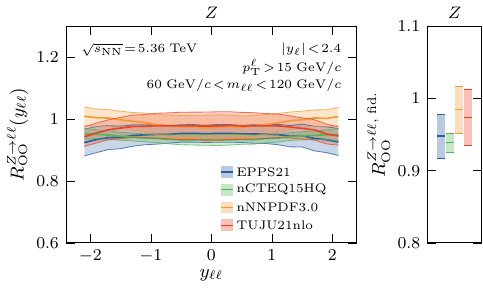}
    \caption{Lepton-pair-rapidity differential and fiducial inclusive \RpO (top) and \ROO (bottom) for $Z$ production. The calculations are performed at NLO with the MCFM code~\cite{MCFM}. The shown uncertainty bands denote the nPDF uncertainties at \SI{68}{\percent} confidence level.}
    \label{fig:ZBosonRatios}
\end{figure}

Figure \ref{fig:ZBosonRatios} shows the Z-boson nuclear modification factors \RpO (top row) and \ROO (bottom row) as a function of the dilepton rapidity $y_{ll}$ (left side panels) and for fiducial inclusive cross sections (right side panels). Since these ratios are less affected by the isospin effects, we now see more directly the impact of the nuclear modifications of parton distributions. For \RpO, we see a clear transition from anti-shadowing dominated region at negative rapidities to shadowing dominated region at positive rapidities. For \ROO, there is a mixture of the two effects at all values of $y_{ll}$ within the considered acceptance, resulting in a relatively flat behaviour as a function of rapidity.

These results highlight the opportunity of constraining the oxygen nPDFs with $W^\pm$ and $Z$ boson data, in particular with the mid-forward production in pO collisions, where we find the largest differences in predictions from different nPDF sets. The $W^\pm$ and $Z$ boson production also serve as an important cross check for nPDF constraints from hadronic observables in pO collisions, such as the hadron production considered in Sec.~\ref{sec:hadronproduction}, but cannot fully replace the latter due to different kinematical and flavour dependence.

\begin{table}
    \centering
    \begin{tabular}{lc|lccc}
    \toprule
        & lumi & & xsec & est.\ num. & proj. \vspace{-0.1cm} \\
        & $[\SI{}{nb^{-1}}]$ & & $[\SI{}{nb}]$ & events & stat.\ unc. \\
        \midrule
        \parbox[t]{3mm}{\multirow{3}{*}{pO}}
        &      &$W^+$ & 59.0 & 2950 & \SI{1.8}{\percent} \\
        & 50.0 &$W^-$ & 46.3 & 2315 & \SI{2.1}{\percent} \\
        &      &$Z$   & 10.2 &  510 & \SI{4.4}{\percent} \\
        \midrule
        \parbox[t]{3mm}{\multirow{3}{*}{OO}}
        &      &$W^+$ & 529.7 & 5297 & \SI{1.4}{\percent} \\   
        & 10.0 &$W^-$ & 443.7 & 4437 & \SI{1.5}{\percent} \\
        &      &$Z$   &  95.4 &  954 & \SI{3.2}{\percent} \\
    \bottomrule
    \end{tabular}
    \caption{Estimated statistics and projected statistical uncertainties for $W^\pm$ and $Z$ boson production with example luminosities mimicking those delivered to CMS during the 2025 light-ion campaign. The total fiducial cross sections are estimated by taking a simple average of the predictions from the four different nPDFs studied in this work. The impact of efficiency corrections is not taken into account in the statistical estimates.}
    \label{tab:WZtable}
\end{table}

To study the feasibility of these measurements, we take the total fiducial cross sections as an average of the predictions from the four different nPDFs studied in this work, and multiply them with luminosities reflecting approximately those delivered to CMS in 2025 to obtain an estimate on the number of events. The obtained numbers and the corresponding projected statistical uncertainties are shown in Tab.~\ref{tab:WZtable}. These estimates do not take into account any efficiency corrections and their impact on the obtainable statistics. In any case, we see that the delivered luminosities are certainly sufficient for total fiducial measurements with statistical uncertainties in the range where new constraints for the nPDFs can be expected. These measurements would still benefit from more luminosity, which would allow for more differential, rapidity dependent measurements, which are difficult to perform with the current estimated number of events.

\subsection{Prompt photon production}
\label{sec:promptphotonproduction}
Prompt photons are produced directly in the hard scattering, dominantly via the LO Compton process ($qg\rightarrow\gamma q$) at LHC energies~\cite{Ichou:2010wc}. 
The measurement of prompt photons is particularly interesting for OO collisions, since the prompt photon does not interact strongly with the potentially formed QGP, and its production is therefore only affected by CNM effects, not by energy loss.
The production cross section is particularly sensitive to the gluon (n)PDF, which is not directly accessible via deep inelastic scattering experiments.
At low $p_{\rm T}$, prompt photons may also be produced in the fragmentation process through quark-photon splittings, which is commonly absorbed in calculations using non-perturbative parton-to-photon fragmentation functions, which are defined for a given factorization scheme at a given factorization scale.
Both theoretical calculations and measurements of prompt photon production commonly employ an isolation requirement, in order to suppress this fragmentation contribution, as well as experimental background from electromagnetic decays (mainly $\pi^0\rightarrow\gamma\gamma$).
This is done by requiring that the energy in the vicinity of the photon is below a given threshold.
Isolated prompt photon production has been measured in pp \cite{ATLAS:2016fta,CMS:2018qao,ALICE:2024kgy}, \pPb \cite{ALICEisophotonpPb,ATLAS:2019ery}, and \PbPb \cite{ALICE:2024yvg,ATLAS:2015rlt,CMS:2012oiv} collisions at LHC energies. 
The measured nuclear modification factors in \pPb and \PbPb collisions are compatible with pQCD calculations at NLO incorporating CNM effects, while the experimental results of absolute pp cross sections hint at the need for NNLO precision for calculations of the prompt photon cross section \cite{ATLAS:2016fta}.

\begin{figure}[t]
    \centering
    \includegraphics[width=\linewidth]{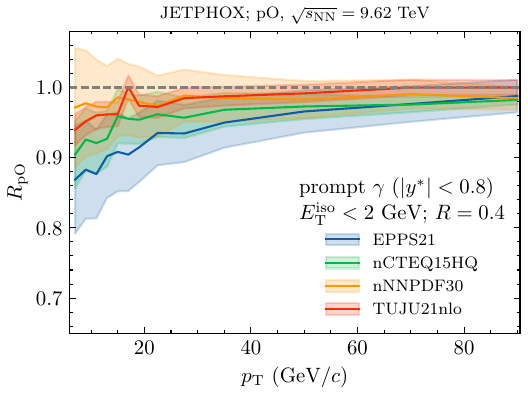}\\
    \includegraphics[width=\linewidth]{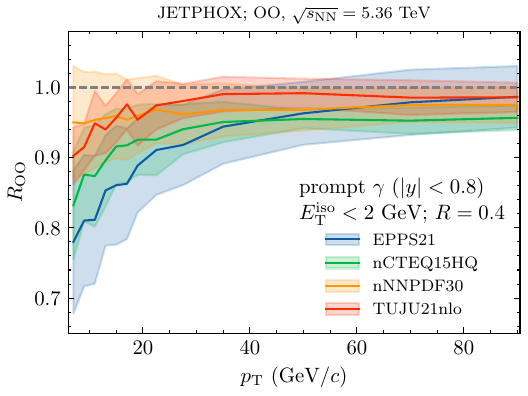}
    \caption{Nuclear modification factors \RpO (top) and \ROO (bottom) for isolated prompt photon production. Shaded bands denote the corresponding nPDF uncertainties at 68\% confidence level.}
    \label{fig:promptphotonROO}
\end{figure}
In this work, we calculate the isolated prompt photon production cross section at mid-rapidity $|y^*|<0.8$ in the centre-of-mass frame using the JETPHOX \cite{jetphox} program at NLO.
In line with a recent ALICE measurement \cite{ALICEisophotonpPb}, we apply a fixed-cone isolation, requiring that the transverse energy $E_{\rm T}^{\rm iso}$ in the cone of radius $R=0.4$ is below $\SI{2}{GeV}$.
The pQCD calculation is performed in pp collisions at $\sqrt{s}=\SI{5.36}{TeV}$, as well as for pO and OO collisions at $\snn=\SI{9.62}{TeV}$ and \SI{5.36}{TeV}, respectively.
As for the other calculations, the used nPDFs are given in Tab. \ref{tab:pdftable} and the nuclear modification factor is calculated according to Eq. \ref{eq:raa_minbias}.
Prompt photons produced in the fragmentation process that survive the isolation requirement are incorporated using the BFG II \cite{BFG2} parton-to-photon fragmentation function, and all scales are chosen to coincide with the photon transverse momentum.

Figure \ref{fig:promptphotonROO} shows the nuclear modification factor of isolated prompt photon production for pO (top) and OO collisions (bottom).
While the scale uncertainties were found to be negligible on the ratios, the nPDF uncertainties are sizeable and denoted by shaded bands.
As currently JETPHOX does not have the ability to use precomputed grids to calculate cross sections for all nPDF members, calculations of the nPDF uncertainties are computationally costly.
The relative nPDF uncertainties are therefore only calculated at LO precision. However, we confirmed that the relative uncertainties obtained at NLO are identical.
The central values are calculated at NLO precision.
As for the inclusive hadron production, agreement with unity is observed for high $p_{\rm T}$, however, significant suppression of up to \SI{20}{\percent} is observed at low $p_{\rm T}$ due to gluon shadowing.
Overall, the suppression is about \SI{10}{\percent} larger than the one observed for hadron production.
This is expected, as a prompt photon at a given $p_{\rm T}$ probes, on average, a lower $x$ than a hadron at the same transverse momentum, since the hadron only carries a small momentum fraction $z$ of the outgoing parton, and in prompt photon production the isospin effects can also lead to an additional suppression with respect to the pp reference~\cite{Arleo:2011gc}.

\section{Hadron to electroweak-boson ratios}
\label{sec:hadron_over_ewboson}

In this section, we consider the possibility of partially cancelling the CNM effects in hadron nuclear modification factor by taking ratios with respective factors for electroweak bosons. As discussed in the previous section, the electroweak bosons and their decay leptons do not interact with the created hot medium and therefore the hot-medium effects from the hadron production in nucleus-nucleus collisions persist in these ratios. However, nPDF effects can at least partially cancel if the probed momentum fractions in the hadron and electroweak-boson production are close enough. This could potentially lead to a more precise theoretical CNM baseline.

\subsection{$Z$-boson normalised hadron \ROO}

One particular example of a hadron over electroweak-boson ratio is the $Z$-boson normalised hadron \ROO,
\begin{equation}
    R^{h^\pm/Z}_\mathrm{OO}(p_{\rm T}) = \frac{R^{h^\pm}_\mathrm{OO}(p_{\rm T})}{R^{Z \rightarrow \ell^+\ell^-,\;\mathrm{fid.}}_\mathrm{OO}} = \frac{\sigma^{Z,\;\mathrm{fid.}}_\mathrm{pp}}{\sigma^{Z,\;\mathrm{fid.}}_\mathrm{OO}} \, \frac{\frac{{\rm d}\sigma^{h^\pm}_\mathrm{OO}}{{\rm d}p_{\rm T}}}{\frac{{\rm d}\sigma^{h^\pm}_\mathrm{pp}}{{\rm d}p_{\rm T}}}. \label{eq:hperZ}
\end{equation}
This ratio was considered previously in Ref.~\cite{Huss:2020dwe} in the context of jet over $Z$-boson production with the aim of cancelling luminosity uncertainties and, potentially, nPDF uncertainties.

\begin{figure}[t]
    \centering
    \includegraphics[width=1.0\linewidth]{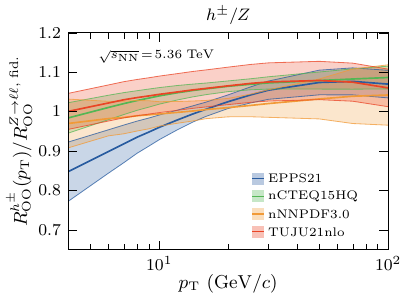}
    \caption{The charged-hadron nuclear modification factor normalised with the $Z$-boson total fiducial \ROO. The hadron and Z-boson acceptance cuts correspond to those used in Secs.~\ref{sec:hadronproduction} and~\ref{sec:ewbosonproduction}.}
    \label{fig:hadronZRatio}
\end{figure}

However, it turns out that nPDF cancellation is not very effective.
We note that for the hadron production in Sec.~\ref{sec:hadronproduction}, the ordering of the central predictions from different nPDFs for the \ROO (which also varies as a function of $p_{\rm T}$) is somewhat different than what is seen in the total fiducial $Z$-boson \ROO. One can thus expect that normalising the hadron nuclear modification factor with the Z boson cross-section ratio is not an effective way to cancel nPDF dependence, as was found to be the case also in Ref.~\cite{Huss:2020dwe} for jet production.

\begin{figure*}[t]
    \centering
    \includegraphics[width=0.49\linewidth]{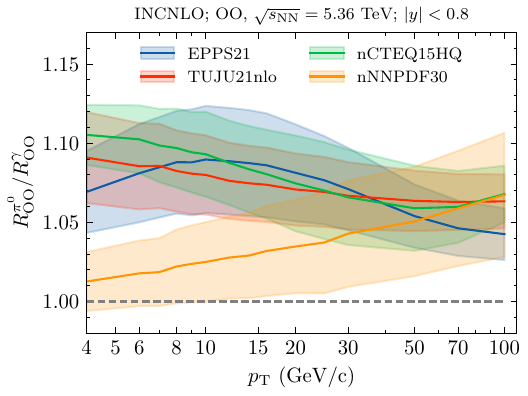}
    \includegraphics[width=0.49\linewidth]{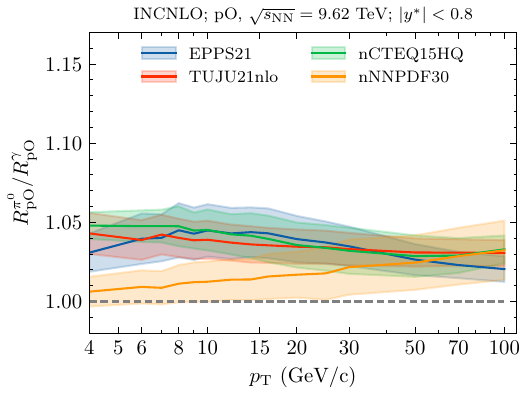}
    \caption{Double ratio $\ROO^{\pi^0/\gamma}(p_{\rm T})$ (left)  and $\RpO^{\pi^0/\gamma}(p_{\rm T})$ (right) of the neutral pion nuclear modification factor with respect to the prompt photon nuclear modification factor. The calculation is performed using the INCNLO program, using the BKK and BFGII fragmentation functions for $\pi^0$ and photon production, respectively. The shaded bands denote the nPDF uncertainties at \SI{68}{\percent} confidence level, where cancellation of nPDF uncertainties for $\pi^0$ and prompt photon production were taken into account.}
    \label{fig:pi0gammaratio}
\end{figure*}

This is indeed verified in Fig.~\ref{fig:hadronZRatio}, where we show the ratio in Eq.~\eqref{eq:hperZ} with the acceptance cuts introduced in Secs.~\ref{sec:hadronproduction} and~\ref{sec:ewbosonproduction}. The total variation and uncertainties of the nPDF baselines are somewhat reduced compared to the hadron \ROO, but the full nPDF variation envelope can still exceed $\pm\SI{10}{\percent}$ uncertainty at the lowest values of hadron $p_{\rm T}$.
This poor nPDF cancellation can be attributed to the hadron and $Z$ production being predominantly sensitive to flavour combinations different from each other, as well as differences in the probed $x$ and $Q^2$ values.

\subsection{Prompt photon normalised pion \ROO}
Next we consider the normalisation of the $\pi^0$ nuclear modification factor with the prompt photon production \ROO~\cite{Reygers:2004hab}, i.e.
\begin{equation}
    R_{\rm OO}^{\pi^0/\gamma}(p_{\rm T}) = \frac{R_{\rm OO}^{\pi^0}(p_{\rm T})}{R_{\rm OO}^{\gamma}(p_{\rm T})}.
    \label{eq:pi0promptgammaroo}
\end{equation}
This ratio has a few experimental and theoretical advantages with respect to the normalisation using $Z$ bosons outlined in the previous section.
Since prompt photon production involves the $qg\rightarrow\gamma q$ channel already at LO, it probes more directly the initial state gluons compared to the previous case where the $Z$ boson is mainly sensitive to the valence- and sea quark content of the projectiles, and therefore the $\pi^0/\gamma$ ratio should allow for a higher degree of cancellation of nPDF uncertainties.
Nonetheless, the involvement of the electromagnetic coupling in the $qg\rightarrow\gamma q$ and $q\bar{q}\rightarrow\gamma g$ channels means an enhanced sensitivity to up quarks, which can lead to imperfect nPDFs cancellation and the $\ROO^{\pi^0/\gamma}$ ratio to deviate from unity simply by isospin effects (cf.\ Ref.~\cite{Arleo:2011gc}).
In addition, as for the $h^{\pm}/Z$ scenario, prompt photon production and $\pi^0$ probe on average different values of $x$~\cite{Helenius:2014qla} at a given $p_{\rm T}$ due to their differing fragmentation, which also impacts the degree of nPDF uncertainty cancellation and absolute value of $\ROO^{\pi^0/\gamma}(p_{\rm T})$.

From an experimental point of view, using $\pi^0$ production, rather than inclusive hadron production, as the numerator offers the possibility for a higher degree of cancellation for experimental uncertainties, as both $\pi^0\rightarrow\gamma\gamma$ and prompt photons can be measured in the same detector subsystem, i.e.\ the electromagnetic calorimeter.
The PHENIX collaboration \cite{PHENIX:2023dxl} has presented a measurement of $\pi^0$ and prompt photon production in \dAu collisions at $\sqrt{s_{\rm NN}}=\SI{200}{GeV}$.
However, rather than the ratio introduced in \Eq{eq:pi0promptgammaroo}, they present the nuclear modification factor as a function of the number of binary collisions $N_{\rm coll}$, where the prompt photons were used to assess centrality biases in peripheral \dAu collisions.
Such an approach may also be useful for future measurements in light-ion collisions when moving from MB collisions discussed in our work towards measurements binned in centrality, which may come with substantial selection biases \cite{Loizides:2017sqq}.
However, as pointed out in Ref.~\cite{Perepelitsa:2024eik}, special care has to be taken when interpreting these results due to the differing $x$ sensitivity of prompt photons and neutral pions.

Figure \ref{fig:pi0gammaratio} (left) shows the double ratio $\ROO^{\pi^0/\gamma}$ introduced in Eq. \ref{eq:pi0promptgammaroo} of the $\pi^0$ and prompt photon nuclear modification factors in minimum-bias collisions.
The calculations have been performed at NLO using the INCNLO \cite{INCNLO1} program, both for $\pi^0$ and prompt photon production.
In contrast to Sec. \ref{sec:promptphotonproduction}, where the JETPHOX program has been used to calculate prompt photon production including the isolation requirement, we found the used numerical precision too limiting for the presented double ratio.
Instead, we calculated the prompt photon production cross section without isolation using the INCNLO program, and verified that the obtained predictions agree with the JETPHOX predictions.
The uncertainty bands in Fig. \ref{fig:pi0gammaratio} are denoted at $\SI{68}{\percent}$ confidence level, and are obtained by simultaneous variation of the nPDF members for the $\pi^0$ and prompt photon calculation.
As expected, due to the sensitivity of prompt photons to slightly lower $x$ with respect to neutral pions, the double ratio is slightly above unity, up to \SI{10}{\percent} at low $p_{\rm T}$.
In any case, we observe significant cancellation of nPDF uncertainties for this observable; the relative nPDF uncertainty of the double ratio is below \SI{2}{\percent} for all nPDFs in the considered $p_{\rm T}$ range. 
This is significantly smaller than the nPDF uncertainties of the individual nuclear modification factors entering the double ratio, which exceed about \SI{10}{\percent} at low-$p_{\rm T}$.
Scale uncertainties are likewise found to be negligible on this double ratio.
However, it appears that nNNPDF3.0 gives a somewhat different $p_{\rm T}$ dependence for this observable, and the total envelope of different nPDF predictions yields an uncertainty of the order of \SI{5}{\percent} at low $p_{\rm T}$.

The good cancellation of nPDF and scale uncertainties makes the double ratio particularly suited for experimental searches for parton energy loss.
The double ratio in pO collisions 
$\RpO^{\pi^0/\gamma}(p_{\rm T}) = R_{\rm pO}^{\pi^0}(p_{\rm T}) / R_{\rm pO}^{\gamma}(p_{\rm T})$ provides a direct way to assess the presence of energy loss in this system.
This double ratio is presented in Fig. \ref{fig:pi0gammaratio} (right). 
As expected, the deviations from unity are smaller than for the $\ROO^{\pi^0/\gamma}$, and a similar cancellation of nPDF uncertainties is observed.
\section{OO to pO hadron ratios}
\label{sec:soo}

\begin{figure*}
\centering
\includegraphics[width=0.49\linewidth]{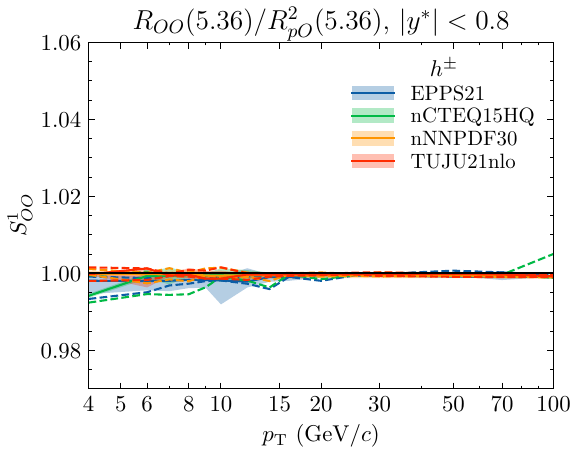}
\includegraphics[width=0.49\linewidth]{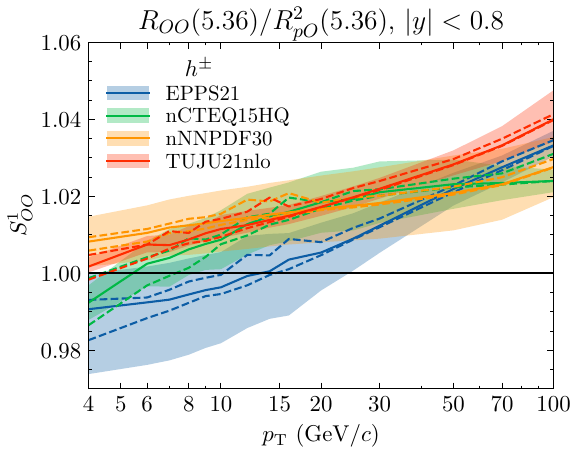}
\caption{Charged hadron $S_{OO}=\ROO/\RpO^2$ ratio with all collisions at the same collision energy $\sqrt{s_{NN}}=5.36\,\text{TeV}$. (Left) All collisions in the same centre-of-mass rapidity window $|y^*|<0.8$ (right) in the same laboratory frame rapidity window $|y|<0.8$. Bands show 68\% nPDF uncertainty intervals. Dashed lines show scale variation envelope. \label{fig:SAA_v1}}
\end{figure*}

In this section we discuss ratios of hadronic cross sections in OO, pp and pO collisions that may help cancel out some of current nPDF uncertainties and increase the significance of energy-loss signals in light-ion collisions. One way to cancel nPDF dependence in oxygen-oxygen hadron nuclear modification factor \ROO is to divide by the square of \RpO. Following Ref.~\cite{ALICE:2015sru} we call this ratio $S_{OO}$,
\begin{equation}
S_{\rm OO} = \frac{\ROO}{\RpO^2}.\label{eq:SAA}
\end{equation}

Because the oxygen nPDF are used twice in the numerator and denominator, the nPDF uncertainties should mostly cancel if OO and pO collisions were performed at the same centre-of-mass energy. During the short run at LHC in 2025, the collision energy between OO and pO was different. In addition, the pO centre-of-mass was boosted by $\Delta y \approx 0.35$ with respect to the lab frame. Therefore, the same experimental acceptance results in different rapidity window for pO system compared to symmetric systems like OO and pp. Furthermore, the pp reference was not taken at pO collision energy. The issue of missing pp reference can be addressed by interpolating from measurements at different energies, or by considering ratios of different collision energies~\cite{Brewer:2021tyv}.

Another important consideration is that, unlike the ratios involving electroweak bosons in Sec.~\ref{sec:hadron_over_ewboson}, the ratio of hadrons in \Eq{eq:SAA} could partially cancel the energy-loss signal. For small energy-loss signal (for a moment neglecting CNM effects) $|1-\RAA|\ll 1$, the energy loss in \pA system would need to be more than twice smaller, $|1-\RpA|\ll |1-\RAA|/2$, for it not to cancel out in \Eq{eq:SAA}. The current measurements of \pPb have not shown an energy-loss signal at the same level as similar multiplicity OO data, which points out that the energy loss might be significantly smaller in these asymmetric systems~\cite{Faraday:2025prr}.  

In this work, we discuss only the cancellation of theoretical uncertainties in CNM baseline. The multi-cross-section ratios, such as \Eq{eq:SAA} might compound the experimental uncertainties. Especially, normalisation or luminosity uncertainties might be difficult to cancel across different collision systems. However, the proper assessment of experimental uncertainties for a specific observable has to be done case by case and is outside the scope of the current paper.

With the caveats outlined above, we consider the several variations of \Eq{eq:SAA} for inclusive charged hadrons. For the computations in this section we include the error sets both of nucleus and proton PDFs (EPPS21, nNNPDF30 and TUJU21), accounting for their correlations in the uncertainty cancellation. As nCTEQ15HQ does not propagate proton baseline uncertainties, we use the central member of CT18NLO in the ratios. In addition, we will compute pO cross sections in two rapidity windows. The symmetric $|y^*|<0.8$ range as for pp and OO collisions, which can be only selected experimentally if the acceptance window is large enough and $-1.15<y^*<0.45$, which corresponds to pO system boosted by $\Delta y=0.35$ in the positive rapidity direction in the lab frame (negative boost from the centre-of-mass-frame viewpoint).

\subsection{\RpO at 5.36 TeV}

\begin{figure}[t]
\includegraphics[width=\linewidth]{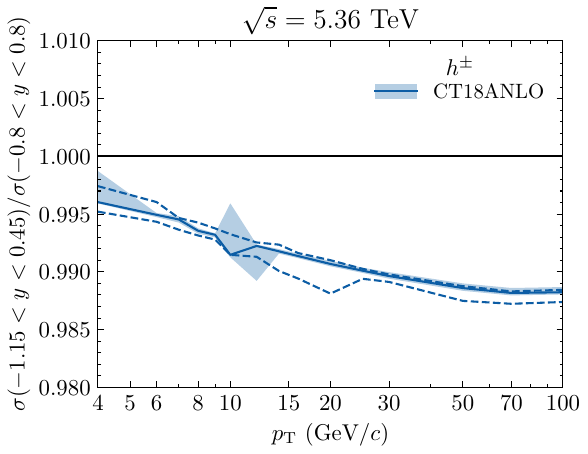}
    \caption{Ratio of charged-hadron cross section in pp at the same collision energy, but different rapidity windows.}
    \label{fig:ppasymmetric}
\end{figure}

\begin{figure*}[t]
\centering
\includegraphics[width=0.49\linewidth]{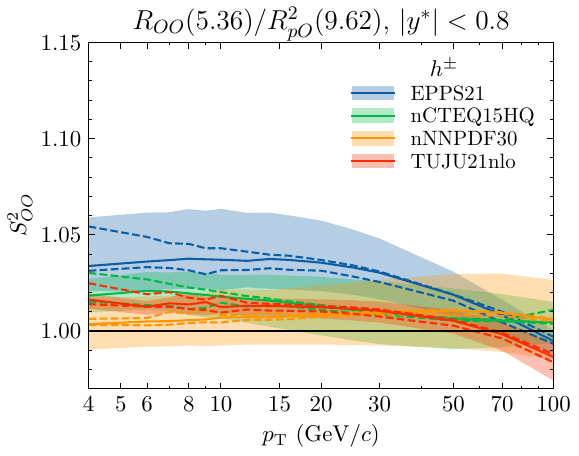}
\includegraphics[width=0.49\linewidth]{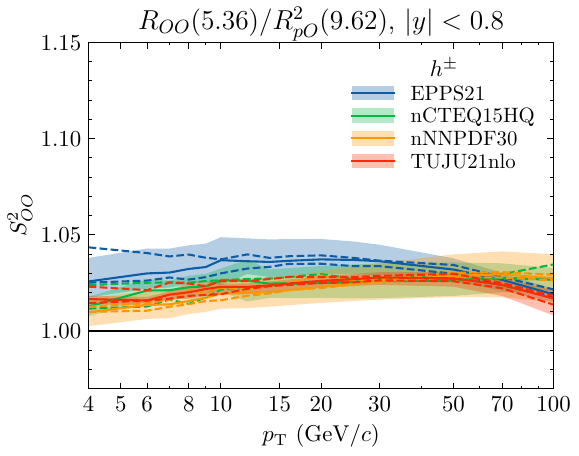}
\caption{Same as \Fig{fig:SAA_v1}, but with OO and pO at different collision energies. (Left) All collision in the same centre-of-mass rapidity window $|y^*|<0.8$ (right) in the same laboratory frame rapidity window $|y|<0.8$. Bands show 68\% nPDF uncertainty intervals. Dashed lines show scale variation envelope.\label{fig:SAA_v2}}
\end{figure*}

First, we consider the case of OO and pO collisions at the same collision energy.
This is the ideal case where we would expect the maximum uncertainty cancellation. The ratio of six cross sections simplifies to
\begin{equation}
S^1_{\rm OO} = \frac{\ROO(5.36)}{\RpO(5.36)^2} =\frac{\frac{d\sigma_{OO}(5.36)}{dp_T}\frac{d\sigma_{pp}(5.36)}{dp_T}}{\frac{d\sigma_{pO}(5.36)}{dp_T}\frac{d\sigma_{pO}(5.36)}{dp_T}}.\label{eq:SAA_v1}
\end{equation}
We label this version of \Eq{eq:SAA} as $S^1_{\rm OO}$ to distinguish it from other variations of $S_{\rm OO}$, which we will consider later.

Results are shown in \Fig{fig:SAA_v1}, where the shaded bands show the resulting 68\% confidence interval, while dashed lines indicate the spread of 15-point scale variation.
The left panel of \Fig{fig:SAA_v1} shows the results when all collisions are measured in the same centre-of-mass rapidity window $|y^*|<0.8$. In this case the nPDF uncertainty cancels to sub-percent level and we become sensitive to 0.1\% numerical precision of the numerical integration. On the right panel, we show the case of where pO is measured in the same lab-frame rapidity window $|y|<0.8$, i.e.\ $-1.15<y^*_{pO}<0.45$.
We observe excellent cancellation of nPDF uncertainties to the percent level, small scale uncertainty and a positive slope in the ratio. Importantly, the spread of different nPDFs is much reduced compared to \ROO in \Fig{fig:hadronROOandRpO}.

For the results shown in \Fig{fig:SAA_v1} we kept the reference pp cross section for pO system in the same rapidity range $|y|<0.8$. In \Fig{fig:ppasymmetric} we show results for the ratio of pp cross section in two rapidity windows using CT18NLO PDFs (results for other PDFs are identical). The PDF and scale uncertainties are at sub-percent level and overall ratio is slightly below unity.
 Therefore, the shift in pp reference for pO would slightly modify the overall ratio, but not the uncertainty bands seen in \Fig{fig:SAA_v1}.

\subsection{\RpO at 9.62 TeV}

Next we consider the case of OO collisions at $\sqrt{s_{\rm NN}}=5.36\,\text{TeV}$ and pO at $\sqrt{s_{\rm NN}}=9.62\,\text{TeV}$.
We also assume the existence of a (potentially interpolated)  pp reference for pO collisions at $\sqrt{s}=9.62\,\text{TeV}$.
Then \Eq{eq:SAA} becomes
\begin{equation}\label{eq:SAA_v2}
S^2_{\rm OO} = \frac{\ROO(5.36)}{\RpO(9.62)^2} =\frac{\frac{d\sigma_{\rm OO}(5.36)}{dp_T}\frac{d\sigma_{\rm pp}(9.62)}{dp_T}\frac{d\sigma_{\rm pp}(9.62)}{dp_T}}{\frac{d\sigma_{\rm pO}(9.62)}{dp_T}\frac{d\sigma_{\rm pO}(9.62)}{dp_T}\frac{d\sigma_{\rm pp}(5.36)}{dp_T}}.
\end{equation}
In \Fig{fig:SAA_v2} we show the NLO computation of this ratio in centre-of-mass (left) and lab frames (right). We observe good cancellation of nPDF uncertainties down to a couple of percents. The ratio is approximately flat, but slightly above unity.
The uncertainty cancellation is somewhat better in the lab frame (right) than in the centre-of-mass frame (left). We expect this to happen due to the higher collision energy in pO pushing the probed nPDFs to smaller $x$, but the rapidity shift conversely leads to higher nuclear $x$ being probed, and the two opposite effects partially compensate for each other in the case of equal lab-frame acceptance measurement.
This is different to what has been shown in Fig~\ref{fig:SAA_v1} (right), where only the latter of these two effects is present.

Compared to \ROO in \Fig{fig:hadronROOandRpO}, the spread of CNM baselines is significantly reduced.
If the pp reference at 9.62 TeV could be measured or reliably interpolated, this ratio might serve as a precise theoretical baseline for energy-loss signals in light-ion collisions. However, as stated also above, multi-cross-section ratios might also compound experimental uncertainties, especially the normalisation uncertainty arising from luminosity determination for each collision system.

We would like to emphasize the importance of documenting the exact interpolation procedure in the construction of pp reference. This would allow one to perform the same interpolation on theoretically computed spectra and their error sets, resulting in well-defined theoretical uncertainties in \Eq{eq:SAA_v2} even with an interpolated reference.

\begin{figure*}
\centering
\includegraphics[width=0.49\linewidth]{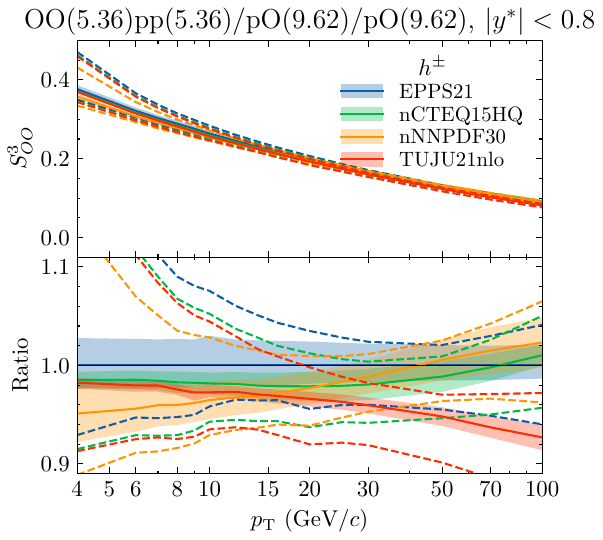}
\includegraphics[width=0.49\linewidth]{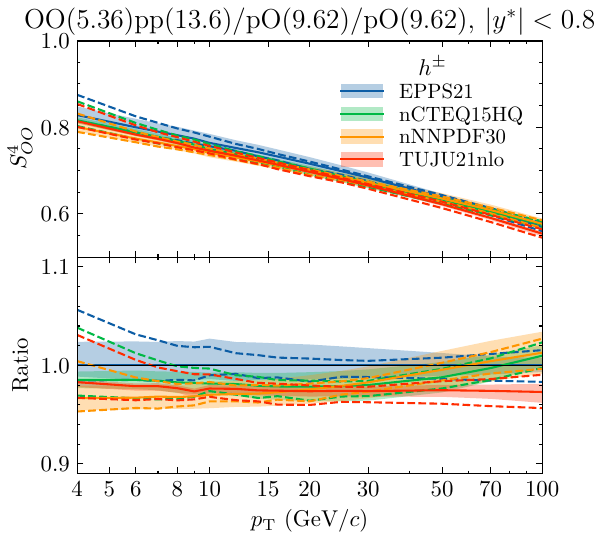}
\caption{
Charged hadron $S_{OO}=\ROO/\RpO^2$ ratio with one pp reference for pO collisions at (left) $\sqrt{s}=5.36\,\text{TeV}$ 
(right) $\sqrt{s}=13.6\,\text{TeV}$.
All collisions are in the same centre-of-mass rapidity window $|y^*|<0.8$. Bands show 68\% nPDF uncertainty intervals. Dashed lines show scale variation envelope. \label{fig:SAA_v3v4}}
\end{figure*}

\subsection{Mixed-energy \RpO}

It is possible to avoid the use of interpolated pp reference at pO collision energy, by using a mixed-energy \RpO. Namely, one can use a pp reference at some other measured collision energies. It is natural to consider the 5.36 TeV pp reference, which partially cancels in the ratio and results in a four cross-section ratio
\begin{equation}
S_{\rm OO}^3 =  \frac{\frac{d\sigma_{\rm OO}(5.36)}{dp_T}\frac{d\sigma_{\rm pp}(5.36)}{dp_T}}{\frac{d\sigma_{\rm pO}(9.62)}{dp_T}\frac{d\sigma_{\rm pO}(9.62)}{dp_T}}.\label{eq:SAA_v3}
\end{equation}
The perturbative cross section increases with collision energy and becomes harder. Therefore, the ratio above is below unity and decreasing with increasing momentum. Results are shown in \Fig{fig:SAA_v3v4}(left). In the top panel, we show the absolute ratio, while in the lower panels we present results normalised by the central EPPS21 line.
In this case we show results only for the centre-of-mass rapidity window $|y^*|<0.8$.

We observe that the nPDF uncertainty cancellation is slightly worse than in \Fig{fig:SAA_v2}.
Strikingly, the scale variation band is significantly larger than nPDF uncertainties, indicating that this ratio might be sensitive to higher-order corrections. Therefore, this ratio is not suitable as a precision CNM baseline.

Instead, we suggest using pp reference at a high-energy, namely 13.6 TeV. That is, we propose the following ratio
\begin{equation}
S_{\rm OO}^4 =  \frac{\frac{d\sigma_{\rm OO}(5.36)}{dp_T}\frac{d\sigma_{\rm pp}(13.6)}{dp_T}}{\frac{d\sigma_{\rm pO}(9.62)}{dp_T}\frac{d\sigma_{\rm pO}(9.62)}{dp_T}}.\label{eq:SAA_v4}
\end{equation}
Now the pO cross section in the denominator is at approximately mean collision energy of OO and pp systems.  Although the higher pp collision energy does not fully cancel the $p_T$ dependence, as shown in \Fig{fig:SAA_v3v4}(right), the ratio is closer to unity. More importantly, the scale variation envelope is below 5\% in the considered region and comparable to nPDF uncertainties. Therefore, this ratio could be used as a precise energy-loss baseline. 

\subsection{Rapidity symmetrized ratio}

All scenarios so far have been focused on predictions for measurements performed in a symmetric acceptance of $|y|<0.8$. 
However, when moving to asymmetric rapidity intervals, commonly encountered for measurements at forward rapidity, the presented double ratio in \Eq{eq:SAA} needs to be modified to the following symmetric ratio
\begin{equation}
    S_{\rm OO} = \frac{\ROO}{\RpO\times R_{\rm Op}},
    \label{eq:r3mod}
\end{equation}
where \RpO and $R_{\rm Op}$ denote the nuclear modification factor for the two different possible directions of the nucleus beam.
This ratio has been previously used in the measurement of forward $J/\psi$ production in \pPb collisions \cite{ALICE:2015sru}, following theoretical calculations \cite{Vogt:2010aa} that likewise introduce the observable equivalent to \Eq{eq:r3mod}.

\begin{figure}[t]
    \centering
\includegraphics[width=\linewidth]{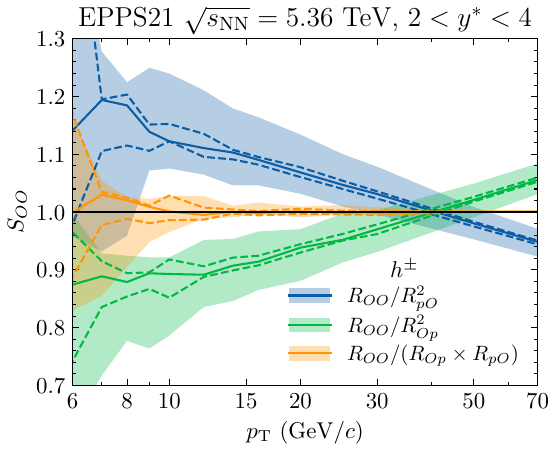}
    \caption{$S_{OO}$ in forward rapidity window for different pO collision orientations. Bands show 68\% EPPS21 nPDF uncertainty intervals. Dashed lines show scale variation envelope. }
    \label{fig:asymmetricreference}
\end{figure}

The symmetrization in \Eq{eq:r3mod} is needed to ensure the optimal cancellation of nPDF uncertainties.
To illustrate this point, we calculate hadron production in pp, pO, Op and OO collisions at the same centre-of-mass energy of $\sqrt{s_{\rm NN}}=\SI{5.36}{TeV}$ at forward rapidity $2<y^*<4$ in the centre-of-mass frame. We do not consider rapidity boosts in order not to complicate the comparison and only use EPPS21 nPDFs.
\Figure{fig:asymmetricreference} shows the three possible ratios which may be constructed considering the two possible directions of the oxygen beam.
Both ratios, $\ROO/\RpO^2$ and $\ROO/R_{\rm Op}^2$, show significant deviations from unity.
This can be understood when considering that a measurement at $2<y^*<4$ in pO collisions is probing oxygen nPDFs at small-$x$, while the same rapidity in Op collisions probes the large-$x$ structure of the nucleus\footnote{For inclusive hadron measurements, the recoiling hadron rapidity is not constrained and introduces smearing in the probed $x$ range, so the arguments here are only heuristic.}. Because nPDFs are typically suppressed at small $x$ (shadowing) and enhanced at larger $x$ (anti-shadowing) (see \Fig{fig:npdfsvsxQ2100GeV}), dividing \ROO by $\RpO^2$ is probing the ratio of CNM effects at large and small $x$. For small $p_T$, this gives a larger than unity ratio. At $p_T\sim 40\,\text{GeV}$ the ratios cross unity, which likely indicates the transition from shadowing to anti-shadowing in pO and from anti-shadowing to EMC effect in Op. The $\ROO/R_{\rm Op}^2$ ratio mirrors the trend reflected at unity.

We note that due to differences in nPDFs between different collaborations (see \Fig{fig:npdfsvsxQ2100GeV}), the $\ROO/\RpO^2$ and $\ROO/R_{\rm Op}^2$ ratios in \Fig{fig:asymmetricreference} look different for different nPDF sets (data not shown). However, the symmetrized ratio \Eq{eq:r3mod} for all nPDFs shows excellent uncertainty cancellation.
This highlights the need for experimental measurements at both forward and backward rapidities for asymmetric proton-nucleus systems. If experimental acceptance is constrained only to forward rapidities, then this requires the run with nucleus in the reversed direction\footnote{We note that pO run at LHC in July 2025 was performed only with proton in the first beam.}.

\section{NeNe to OO hadron ratio}
\label{sec:rnene_roo}
\begin{figure}
\centering
\includegraphics[width=\linewidth]{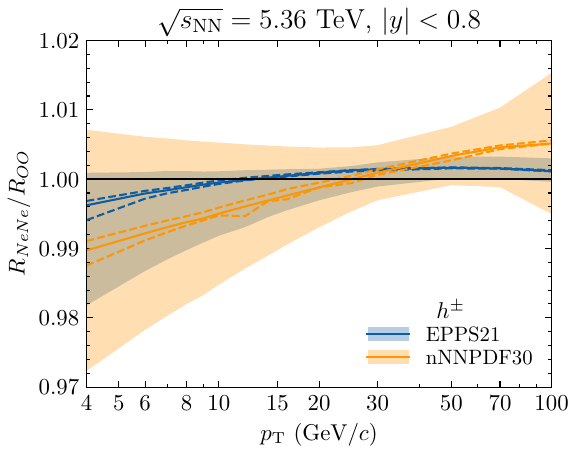}
\caption{The ratio of NeNe and OO inclusive-hadron cross sections. The bands show 68\% nPDF confidence intervals.  Dashed lines show the scale variation envelope. \label{fig:RNeNe_OO}}
\end{figure}

In this section we briefly mention the possibility to consider the ratios of hadron nuclear-modification factor NeNe over OO~~\cite{vanderSchee:2025hoe,Mazeliauskas:2025clt}. At the same collision energy, this completely cancels the need of pp reference resulting in
\begin{equation}
    \frac{\RNeNe}{\ROO} = \frac{\frac{1}{20^2}\frac{d\sigma_{\rm NeNe}(5.36)}{dp_T}}{\frac{1}{16^2}\frac{d\sigma_{\rm OO}(5.36)}{dp_T}}.
\end{equation}
We show pQCD baseline computations for this ratio in \Fig{fig:RNeNe_OO}. The nPDF uncertainties cancel to a couple percent level and there is much better agreement between EPPS21 and nNNPDF30 compared to regular \RNeNe, see \Fig{fig:hadronROOandRpO}. Interestingly, EPPS21 shows stronger cancellation than nNNPDF30, which is likely because for nNNPDF30 nPDFs the $A$ dependence is encoded in a neural network and less constrained than for EPPS21, i.e.\ for EPPS21 oxygen and neon nPDFs are more correlated. In these ratios the scale uncertainty is negligible.

Naturally, in this ratio the energy-loss signal would also largely cancel. If the energy-loss signal scales with $1-\RAA\propto A^{1/3}$~\cite{Faraday:2025prr} then one would expect
\begin{equation}
  1-  \frac{\RNeNe}{\ROO} \approx 0.07\times (1-\ROO).
\end{equation}
For $\ROO\approx 0.7$, this would give $ \frac{\RNeNe}{\ROO}\approx 0.95$. This is consistent with model calculations~\cite{vanderSchee:2025hoe} and could be distinguishable from the precise nPDF baseline if experimental uncertainties could be also cancelled to a great extent. Of course, since no direct constraints for the small-$x$ gluons in neon are expected in the near future, this assumes that the neon nPDFs can be reliably interpolated with constraints from measurements with oxygen and heavier nuclei.

\section{Summary}
\label{sec:summary}

In this work, we have presented a comprehensive set of perturbative QCD baseline predictions for cold nuclear matter (CNM) effects in light-ion collisions at LHC energies. 
Using recent nuclear parton distribution functions (nPDFs), we computed nuclear modification factors for hadronic and electroweak probes, with the aim of supporting ongoing experimental searches for parton energy loss in small systems.
In Sections \ref{sec:lhapdf} through \ref{sec:ewbosonproduction} we quantified the sizeable nPDF uncertainties arising from the lack of light-ion collider data in the global analyses and the weakly constrained $A$-dependence of CNM effects and showed their impact on the traditional nuclear modification factors. 

Based on our calculations and earlier similar findings, we argue that precise knowledge of these CNM effects is essential to interpret the nuclear modification factors of hard probes in light-ion collisions, which are just starting to become available now.
To this end, we discussed the potential for new nPDFs constraints from measurements in the light-ion collisions with observables arguably free from hot QGP effects.
Furthermore, to aid the searches for the onset of parton energy loss in small collision systems, we study various observables for which the CNM effects largely cancel.

In Section~\ref{sec:hadron_over_ewboson}, we explored multi-cross-section ratios involving hadrons and electroweak bosons. The hadron-to-$Z$ ratio in OO collisions~(\Fig{fig:hadronZRatio}) exhibits only a partial cancellation of nPDF uncertainties, mainly due to the sensitivity to different parton flavor combinations and $x$ regions in the two probes.
In contrast, the double ratio of neutral pion to prompt photon nuclear modification factors~(\Fig{fig:pi0gammaratio}) achieves a strong cancellation of nPDF uncertainties, resulting in an observable with reduced CNM baseline dependence that remains sensitive to potential energy-loss effects. 

In Section~\ref{sec:soo} we studied the cancellation of CNM effects using hadron nuclear modification factor in pO collisions. In particular the double ratio $S^1_{\rm OO} = \ROO(\SI{5.36}{TeV})/\RpO(\SI{5.36}{TeV})^2$ (Fig.~\ref{fig:SAA_v1}) achieves excellent cancellation of nPDF uncertainties.
As no pO measurement at $\sqrt{s_{\rm NN}}=\SI{5.36}{TeV}$ is available, and also the pO system is boosted by $\Delta y\approx0.35$, several variations of this ratio depending on experimental constraints, are discussed.
The ratio $\ROO(\SI{5.36}{TeV})/\RpO(\SI{9.62}{TeV})^2$ (Fig.~\ref{fig:SAA_v2}) offers a very good uncertainty cancellation.
The construction of the double ratio with a mixed energy \RpO can lead to sizeable scale uncertainties.
Nonetheless, we showed that the double ratio ${\rm OO}(\SI{5.36}{TeV}){\rm pp}(\SI{13.6}{TeV})/{\rm pO}(\SI{9.62}{TeV})^2$ gives a reasonably good cancellation of both the scale and nPDF uncertainties and is found to be preferable over using a pp reference at $\sqrt{s}=\SI{5.36}{TeV}$ for this type of ratio (Fig.~\ref{fig:SAA_v3v4}).
For measurements at forward rapidities, it is also important to use a rapidity-symmetrised construction for the nPDF cancellation~(\Fig{fig:asymmetricreference}).
Finally, in Section~\ref{sec:rnene_roo}, the ratio of hadron production in NeNe over OO collisions~(Fig. \ref{fig:RNeNe_OO}) shows that correlated CNM effects between the two systems lead to significant cancellation of nPDF uncertainties.

In summary, the results presented in this work provide a unified set of CNM baselines and precision observables for light-ion collisions. 
We demonstrate that current nPDF uncertainties constitute a major limitation for quantitative studies of parton energy loss in small systems, but also identify a set of experimentally accessible ratios in which these uncertainties largely cancel. 
These observables offer promising avenues for future measurements aimed at establishing or constraining energy loss in light-ion collisions.

\section*{Acknowledgments}
FJ thanks Maurice Coquet and Michael Klasen for the interesting and useful discussions.
AM thanks Tanjona Rabemananjara from NNPDF collaboration for providing neon nPDF grids.
CL acknowledges financial support by the U.S.\ Department of Energy, Office of Science, Office of Nuclear Physics, under contract number DE-SC0005131.
AM was supported by the DFG through the Emmy Noether Programme (project number 496831614) and CRC 1225 ISOQUANT (project number 27381115).

\section*{Datasets}
All numerical results presented in this work -- including both the plotted observables and the underlying quantities used for the derived ratios -- are available in Ref.~\cite{GitLabCNMData}.

\bibliographystyle{utphys}
\bibliography{biblio}

\end{document}